\documentclass[a4paper,preprintnumbers,floatfix,superscriptaddress,pra,twocolumn,showpacs,notitlepage,longbibliography]{revtex4-1}
\usepackage[T1]{fontenc}
\usepackage[sc,osf]{mathpazo}
\usepackage{amsmath, amsthm, amssymb,amsfonts,mathbbol,amstext}
\usepackage{graphicx}
\usepackage{dcolumn}
\usepackage{bm}
\usepackage{bbm}
\usepackage{hyperref}
\usepackage{mathtools}
\usepackage{comment}
\usepackage{color}
\usepackage{multirow}
\usepackage{graphicx}
\usepackage[dvipsnames]{xcolor}
\usepackage[normalem]{ulem}

\newtheorem{Proposition}{Proposition}

\newcommand{\Tr}{{\mathrm{Tr}}}

\def\1{\mathbf{1}}
\def\0{\mathbf{0}}

\newcommand{\ket}[1]{| #1 \rangle}
\newcommand{\bra}[1]{\langle #1 |}

\providecommand{\openone}{\mathbbm{1}}
\renewcommand{\rho}{\varrho}

\newcommand{\processnext}[1]{%
  \ifx\listfinish#1\empty\else\listact{#1}\expandafter\processnext\fi}

\makeatletter
\newsavebox{\@brx}
\newcommand{\llangle}[1][]{\savebox{\@brx}{\(\m@th{#1\langle}\)}%
  \mathopen{\copy\@brx\kern-0.5\wd\@brx\usebox{\@brx}}}
\newcommand{\rrangle}[1][]{\savebox{\@brx}{\(\m@th{#1\rangle}\)}%
  \mathclose{\copy\@brx\kern-0.5\wd\@brx\usebox{\@brx}}}
\makeatother															%Double bra-ket

\newcommand{\ms}{\hphantom{-}}

{\begin{framed}\begin{small}}
{\end{small}\end{framed}}

\DeclareGraphicsExtensions{.pdf,.png,.jpg}

\makeindex

\begin{document}
\title{Semidefinite tests for quantum network topologies}
\date{\today}

\author{Johan {\AA}berg}
\affiliation{Institute for Theoretical Physics, University of Cologne, Germany}
\author{Ranieri Nery}
\affiliation{International Institute of Physics, Federal University of Rio Grande do Norte, 59070-405 Natal, Brazil}
\author{Cristhiano Duarte}
\affiliation{
 Schmid College of Science and Technology, Chapman University 
  1 University Dr, Orange, CA 92866
}
\author{Rafael Chaves}
\affiliation{International Institute of Physics, Federal University of Rio Grande do Norte, 59070-405 Natal, Brazil}
\affiliation{School of Science and Technology, Federal University of Rio Grande do Norte, 59078-970 Natal, Brazil}

\begin{abstract}
Quantum networks play a major role in long-distance communication, quantum cryptography, clock synchronization, and distributed quantum computing. Generally, these protocols involve many independent sources sharing entanglement among distant parties that, upon measuring their systems, generate correlations across the network. The question of which correlations a given quantum network can give rise to, remains almost uncharted. Here we show that constraints on the observable covariances, previously derived for the classical case, also hold for quantum networks. The network topology yields tests that can be cast as semidefinite programs, thus allowing for the efficient characterization of the correlations in a wide class of quantum networks, as well as systematic derivations of device-independent and experimentally testable witnesses. We obtain such semidefinite tests for fixed measurement settings, as well as parties that independently choose among collections of measurement settings. The applicability of the method is demonstrated for various networks, and compared with previous approaches.
\end{abstract}

\maketitle

A basic scientific goal is the development of causal models explaining observed phenomena. Through the mathematical theory of causality,  empirical data can be turned into a causal hypothesis that can be falsified or refined by new observations \cite{Pearl2009,spirtes2000causation, Kleinberg2013}. Not surprisingly, this causal framework has found many applications, ranging from economics \cite{chen2008causal,AK01} to biology and medicine \cite{friedman2004inferring,KH11} and also quantum physics \cite{Henson2014,Ried15,Chaves2015a,Fritz2016,allen2017quantum,costa2016quantum,barrett2019quantum}. Indeed, Bell's theorem \cite{Bell1964} can be regarded as a particular case of a causal inference problem \cite{Spekkens2015,Chaves2015b}, and the phenomenon generally known as quantum nonlocality shows that  quantum correlations are incompatible with our classical notion of cause and effect~\cite{Brunner2014}.

More recently, in view of steady experimental advances \cite{carvacho2017experimental,saunders2017experimental,Liao2019,sun2019experimental}, understanding the role of causality in quantum networks of growing size and complexity, typically composed of many independent sources of entanglement, has become a topic of particular relevance \cite{branciard2019violation}. On the practical side, quantum correlations can be distributed across the whole network via quantum repeaters \cite{briegel1998quantum}. Fundamentally, new and stronger notions of nonlocality can emerge in such quantum networks \cite{Renou2019_2} and lead to novel quantum information protocols.

In spite of its clear importance, the characterization of correlations in quantum networks remains in its infancy. Even in the simplest case of two distant laboratories sharing quantum states, this characterization turns out to be extremely demanding \cite{NPA1}. The situation for more complex quantum networks, such as the quantum internet \cite{Kimble2008,wehner2018quantum}, and quantum repeaters \cite{briegel1998quantum}, is yet more intractable. Due to the independence of entanglement sources, correlations compatible with a quantum network form a non-convex set \cite{Geiger1999,Chaves2016} that, even in a purely classical case, cannot be easily characterized beyond very small networks \cite{Garcia2005,Lee2017, WSF19}. To circumvent this problem, novel approaches have recently been  proposed to characterize quantum causal structures \cite{Chaves2015a,wolfe2019quantum,Renou2019,gisin2020constraints}. In this context, we consider arbitrary quantum networks where a number of distant parties share quantum states provided by several independent sources, and we introduce a method to characterize the covariances that are compatible with such quantum networks, determining the constraints on correlations arising from the topology of the network irrespective of whether the underlying sources are classical or quantum.
Despite the non-convex nature of the underlying problem, our approach can be implemented via efficient semidefinite programs, which moreover allows for the derivation of device-independent witnesses for the quantum network topology. We demonstrate the applicability of our method in various cases also comparing it with previous approaches.

\begin{figure}
\begin{center}
 \includegraphics[width=0.95\columnwidth]{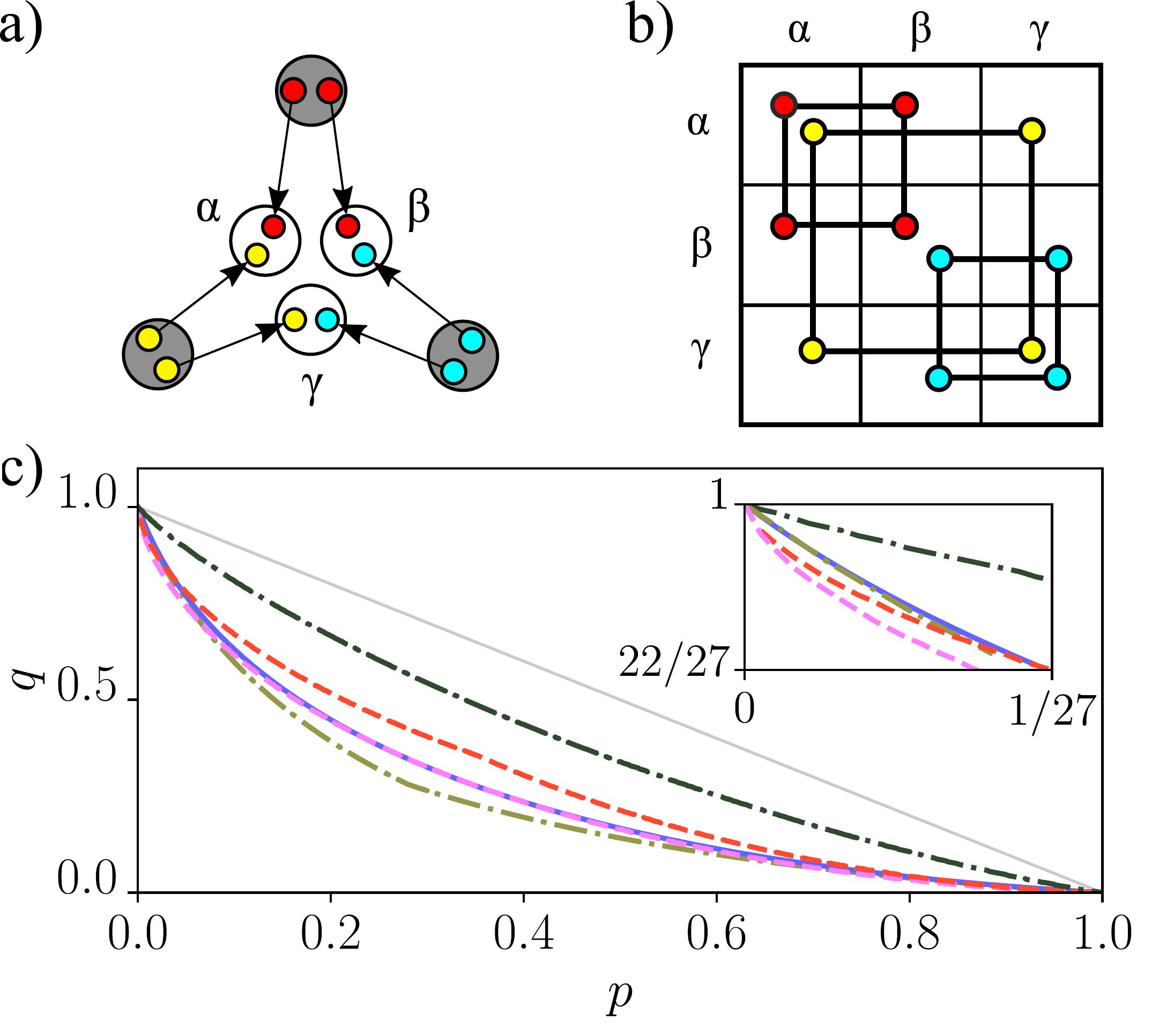}
\end{center} 
\caption{\label{fig:triangle} 
{\bf Triangular quantum network.} \textbf{a)} Triangular quantum DAG with three children $\alpha,\beta,\gamma$ (white disks). Each parent (gray disks) has two children. The parents are  uncorrelated, but distribute (possibly entangled) quantum systems (colored disks) to all of its children, and each child performs a measurement on its systems. 
\textbf{b)} The covariance matrix of the observed random variables can be decomposed into a sum of positive semidefinite blocks; one for each parent, where the support of the block is determined by the children of the parent.
For the triangular DAG, there are three positive semidefinite components (red, yellow, blue), each with bipartite supports. If a given covariance matrix cannot be decomposed in this manner, then the triangular quantum scenario is rejected as an explanation of the observed covariance. 
\textbf{c)} The witness $W_{\mathrm{GHZ}}$ [Eq. \eqref{eq:triangle_witness}] rejects the triangular quantum DAG as an explanation of the distributions $P_{p,q}$  in Eq.~(\ref{eq:dist_pq_tripartite}), establishing the analytic expression $q > p + (4 - \sqrt{1+48\,p})/3$ for the incompatibility region determined by our method. Remarkably,  $W_{\mathrm{GHZ}}$ is optimal for all points above the solid blue curve (verified numerically by comparison with the general semi-definite results in Proposition 1). For comparison we also consider two versions of  the Finner inequality \cite{Renou2019} (red and magenta dashed curves) that differ in terms of local post-processings of the observable variables. First, by using the variables directly without any post-processing, the test rejects all cases above the dashed red curve. Then, by resorting to a numerical optimization over the possible processings, we obtain the magenta curve, that approximates our curve from below and rejects a slightly larger region. Details of the comparison with the Finner inequality are provided in the Supplemental Material \cite{SuppMat}. We consider also the inflation bound (green dot-dashed) recently obtained in \cite{gisin2020constraints} and the entropic bound (black dot-dashed) \cite{henson2014theory,Chaves2015a,weilenmann2017analysing} (see \cite{SuppMat} for more details).}
\end{figure}

We consider typical quantum networks with two types of vertices: i) quantum states distributed among several distant parties, and ii) classical variables standing for the outcomes of measurements performed on such states. These networks can be represented as directed acyclic graphs (DAGs) $G=(V,E)$, consisting of a set of vertices $V$ and directed edges $E$ together with a bipartition of the vertex-set $V$ into  the \emph{latent set} (quantum states) and the \emph{observable set} (measurement outcomes). 
Due to the state-distribution scenario, we here exclusively consider the class of DAGs where all edges are directed from latent vertices to observable vertices, but with no edges within these two groups (see Fig. \ref{fig:triangle}). For this reason, we refer to the elements of the latent set as \emph{parents} $p_n$, and  the elements in the observable set as \emph{children} $c_m$.

Although the quantum state $\rho_{p_n}$, associated to each parent $p_n$,  can be  entangled \emph{internally}, the parents have no correlations between each other, which results in the joint state $\rho = \rho_{p_1}\otimes\cdots \otimes\rho_{p_N}$.  The parents distribute the quantum systems to their children, where child $c_m$ measures an arbitrary positive-operator valued measure \footnote{The values assumed by a POVM are positive semi-definite operators on a Hilbert space and are the most general kind of measurements in quantum theory.} (POVM) $\{A^{(m)}_{x_m}\}_{x_m}$ (see Fig. \ref{fig:triangle}).  For a given graph $G$ with $N$ parents and $M$ children, the joint measurement outcome  $x_1,\ldots,x_M$ thus occurs with a probability 
\begin{equation}
\label{QuantumWithoutInputs}
P(x_1,\ldots,x_M) = \Tr([A^{(1)}_{x_1}\otimes\cdots\otimes A^{(M)}_{x_M}][\rho_{p_1}\otimes\cdots\otimes\rho_{p_N}]).
\end{equation}

The  question is whether a quantum causal structure $G$ can `explain' an observed distribution, in the sense that it can be written as in (\ref{QuantumWithoutInputs}), for \emph{some} choices of states and POVMs that are compatible with $G$.
 Proposition \ref{Prop:DecomposingQuantumNoInputs}, below, provides a method to falsify such causal explanations. For this purpose, we consider a collection of orthogonal Hilbert spaces $\mathcal{V}_1,\ldots,\mathcal{V}_M$, and let $\mathcal{V}:=\bigoplus_{m=1}^{M}\mathcal{V}_m$. To each possible outcome $x_m$ of child $c_m$ we associate a vector $Y^{(m)}_{x_m}\in\mathcal{V}_m$. This type of mappings is often referred to as  \emph{feature maps} \cite{Schoelkopf2002learning}, and can be chosen freely as part of the analysis (see \cite{SuppMat} for discussions). Here, we take the liberty of overloading the notation, and let $Y^{(m)}$ denote both the feature map, and the random vector that results from applying the feature map to the random measurement outcomes $x_m$. By combining all the children, we get the global random vector $Y := \sum_{m=1}^{M}Y^{(m)}$ with elements $Y_{x_1,\ldots,x_M}:= Y^{(1)}_{x_1}+\cdots +Y^{(M)}_{x_M}$.

 Broadly speaking, Proposition \ref{Prop:DecomposingQuantumNoInputs} says that the covariance matrix
$\mathrm{Cov}(Y) = E(YY^{\dagger})-E(Y)E(Y)^{\dagger}$ necessarily 
satisfies a  specific semidefinite decomposition, Eq.~(\ref{vdfbvsdf}), determined by $G$.
This can be viewed as the quantum generalization of a similar result for classical networks in \cite{KelaEtAl20} (see \cite{SuppMat} for a brief summary). 
For intuition concerning the decomposition (\ref{vdfbvsdf}), it can be useful to view $\mathrm{Cov}(Y)$ as a block-matrix with respect to the subspaces $\mathcal{V}_m$, where the corresponding projectors $P_m$ can be used to select  `blocks' $P_m\mathrm{Cov}(Y)P_{m'} = \mathrm{Cov}(Y^{(m)},Y^{(m')})$. Equation  (\ref{Eq.ThmProjectorsCAndR}) forces each $C_n$ to have zero blocks $P_mC_n P_{m'} = 0$ whenever $c_m$ or $c_{m'}$ is not a child of parent $p_n$ . Similarly, (\ref{Eq.ThmProjectorsCAndR}) forces $R$ to be block-diagonal. (See Fig.~\ref{fig:triangle} b,  and an illustration in \cite{SuppMat}.)
\begin{Proposition}
\label{Prop:DecomposingQuantumNoInputs}
Let the distribution $P(x_1,\ldots,x_M)$ be compatible, in the sense of  (\ref{QuantumWithoutInputs}), with a quantum causal structure $G$ with parents $p_1,\ldots,p_N$ and children $c_1,\ldots,c_M$.
 Assume that each child $c_m$  is assigned a feature map $Y^{(m)}$ into a vector space $\mathcal{V}_m$. 
Then there exist operators $R$ and $(C_n)_{n=1}^{N}$ on $\mathcal{V} := \bigoplus_{m=1}^{M}\mathcal{V}_{m}$, such that
 the covariance matrix of $Y= \sum_{m=1}^{M}Y^{(m)}$ satisfies
\begin{eqnarray}
\label{vdfbvsdf}
\mathrm{Cov}(Y) & = &  R + \sum_{n=1}^{N}C_n,\quad R\geq 0,\quad C_n\geq 0,\\
\label{Eq.ThmProjectorsCAndR}
& & P^{(n)}C_nP^{(n)} =C_n,\quad R = \sum_{m=1}^{M}P_mRP_m,
\end{eqnarray}
for the projectors $P^{(n)} := \sum_{m\in \mathcal{C}_n}P_m$, where $\mathcal{C}_n$ denotes the children of parent $p_n$ in the given DAG $G$, and where $P_m$ is the projector onto $\mathcal{V}_m$.
\end{Proposition}
The proof  of  Proposition \ref{Prop:DecomposingQuantumNoInputs} is presented in \cite{SuppMat}.
Since the semidefinite decomposition in Eq.~(\ref{vdfbvsdf}) is a necessary condition, it defines a `semidefinite test' whose failure falsifies $G$ as an explanation of the observed distribution. 
This test is identical to the classical counterpart in  \cite{KelaEtAl20} (see \cite{SuppMat}), 
irrespective of the classical or quantum nature of the underlying sources of correlations.
\begin{figure}
\begin{center}
 \includegraphics[width=0.95\columnwidth]{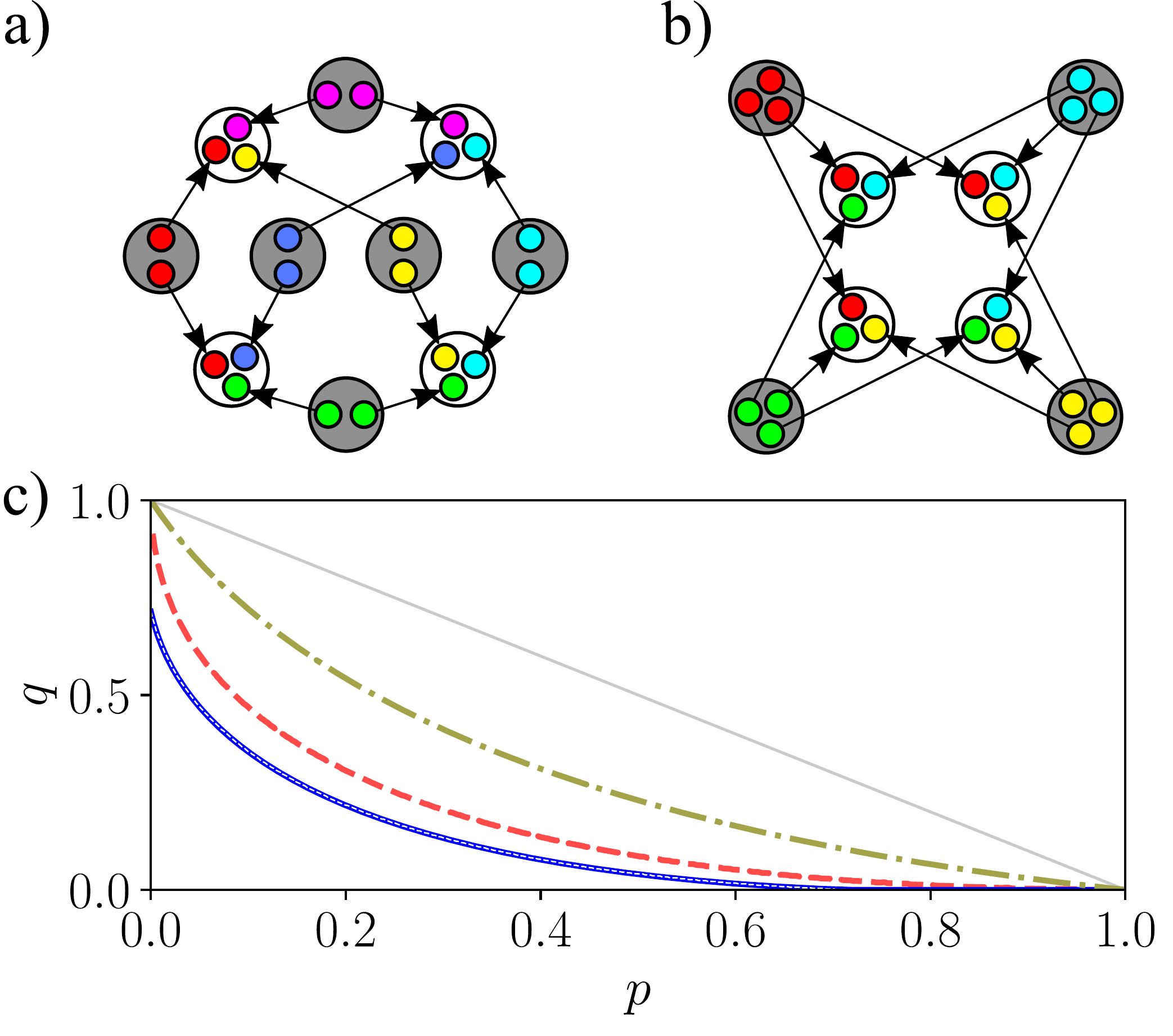}
\end{center} 
\caption{\label{fig:4partite} 
{\bf Tests for 4-partite scenarios.} 
Analogous to Fig.~\ref{fig:triangle}, we consider the family of four-partite distributions $P_{p,q}=p\delta_{0000} + q\delta_{1111} + (1-p-q)(1-\delta_{0000}-\delta_{1111})/14$, with 
$p,q\geq 0$ and $p+q\leq 1$. We test for values of $p$ and $q$ for which the resulting covariance matrix is  compatible with two quantum scenarios: \textbf{a)}  Six parents (grey disks),  each with two children (white disks). \textbf{b)} Four parents, each with three children. \textbf{c)} For all values of $p,q$ above the solid blue curve, our test rejects scenario a) as an explanation of $P_{p,q}$. Above the dot-dashed curve, the test rejects scenario b) as an explanation of $P_{p,q}$.  In scenario a) we moreover compare again with two versions of the Finner test \cite{Renou2019}, with and without optimization over local post-processings of the observable variables. For the latter, we obtain the dashed red curve, while for the former, we obtain a curve that approximately coincides with our curve from above, shown superimposed in light dashed line on top of the solid blue curve. See the Supplemental Material for details \cite{SuppMat}.
}
\end{figure}
The semidefinite test can moreover be cast as a semidefinite program, which is efficiently solved via standard convex optimization tools \cite{SuppMat}, in spite of the  non-convex nature of the underlying problem. From technical point of view, the semidefinite test provides an outer relaxation to the (generally non-convex) set of  distributions that can be explained by $G$. One should note that for applications of the semidefinite test it suffices to know the covariance matrix $\mathrm{Cov}(Y)$, that is, bipartite information. Arguably, this appears advantageous for experimental implementations, since it may be challenging to obtain good estimates of the full distribution of measurement outcomes $P(x_1,\ldots,x_M)$.

One can furthermore derive general purpose witnesses. An equivalent (dual) formulation of \eqref{vdfbvsdf} yields a Hermitian matrix $W$, with the same dimensions as $\mathrm{Cov}(Y)$, such that $\Tr(W\,\mathcal{C}) \leq 0$, for any covariance matrix $\mathcal{C}$ that admits the decomposition in Eq.~\eqref{vdfbvsdf}. 
For example, measuring the observable $\sigma_z$ on each qubit of a GHZ state \cite{greenberger1989going} $\ket{GHZ}=(\sqrt{1/2})\left( \ket{000}+\ket{111}\right)$, generates the distribution  $P_{\mathrm{GHZ}} =\delta_{000}/2+\delta_{111}/2$ with $\delta_{abc}(x_1,x_2,x_3) = \delta_{a,x_1}\delta_{b,x_2}\delta_{c,x_3}$, which has the optimal witness (see \cite{SuppMat})
\begin{equation}
W_{\rm GHZ} \coloneqq \frac{1}{6}\begin{bmatrix}
-1 & \ms 1 & \ms 1 & -1 & \ms 1 & -1 \\
\ms 1 & -1 & -1 & \ms 1 & -1 & \ms 1 \\
\ms 1 & -1 & -1 & \ms 1 & \ms 1 & -1 \\
-1 & \ms 1 & \ms 1 & -1 & -1 & \ms 1 \\
\ms 1 & -1 & \ms 1 & -1 & -1 & \ms 1 \\
-1 & \ms 1 & -1 & \ms 1 & \ms 1 & -1
\end{bmatrix},
\label{eq:triangle_witness}
\end{equation}
where we assume feature maps that assign orthonormal vectors to all outcomes.
For the corresponding covariance matrix $\mathcal{C}_{\mathrm{GHZ}}$, we obtain $\Tr(W\,\mathcal{C}_{\mathrm{GHZ}}) = 0.5$, revealing that the distribution is incompatible with the triangle network (Fig. \ref{fig:triangle}). In \cite{SuppMat} we show the explicit form of the dual problem, as well as generalized forms of the witness for more parties.

To demonstrate the applicability of the semidefinite test, we apply it to the family of distributions
\begin{equation}
\label{eq:dist_pq_tripartite}
P_{p,q}= p \delta_{000}+q \delta_{111}+\frac{1}{6}(1-p-q)(1-\delta_{000}-\delta_{111}),
\end{equation}
where $p,\,q\geq 0$ and $p+q\leq 1$.
In Fig. \ref{fig:triangle}, we compare our method with other constraints proven to hold for quantum/non-signalling correlations: the Finner inequality \cite{Renou2019}, the inflation bound recently obtained in \cite{gisin2020constraints} and the entropic bound \cite{henson2014theory,Chaves2015a,weilenmann2017analysing}. The Finner inequality is the best method in regions close to a deterministic distribution, while the inequality in \cite{gisin2020constraints} detects the incompatibility of $P_{p,q}$ with the triangular quantum DAG in a slightly larger region of parameters as compared with our method. Even though the original Finner inequality \cite{finner1992generalization} is valid for any classical network, the quantum generalization of the Finner inequality, and of the inflation constraints, were proved only for quantum networks where each source is connected with at most two parties. Fig.~\ref{fig:4partite} illustrates the advantage of our method, in that it allows for connections to any number of parties. Furthermore, the Finner and inflation bounds are computationally much more costly: Finner \cite{Renou2019} requires optimizing over post-processing functions while the inflation \cite{gisin2020constraints} requires a Fourier-Motzkin elimination (of double exponential complexity) and in fact has only be done for the triangle network. In turn, our method relies on a computationally efficient semi-definite program and can be used to derive analytical constraints for networks of any size (see \cite{SuppMat} for more details).

\begin{figure}
\begin{center}
 \includegraphics[width=0.95\columnwidth]{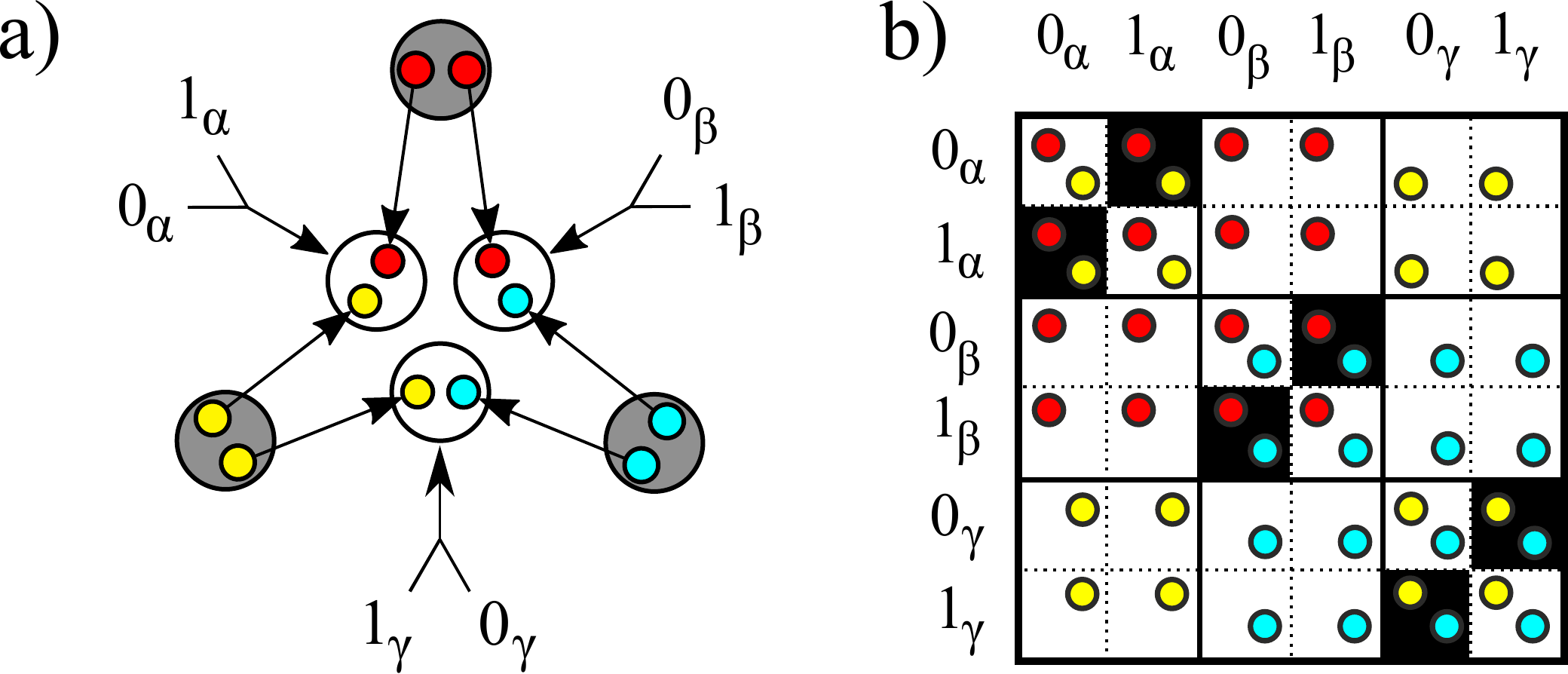}
\end{center} 
\caption{\label{fig:triangleWithChoice} 
 {\bf Triangular quantum network with inputs.}
 \textbf{a)} For the same triangular quantum DAG as in Fig.~\ref{fig:triangle}, we allow each child $m= \alpha,\beta,\gamma$ to choose among two different measurement settings (POVMs). Child $\alpha$ chooses between settings  $0_{\alpha}$ and $1_{\alpha}$, and analogous for child $\beta$ and $\gamma$.
\textbf{b)} For all possible measurement choices we form a covariance matrix of the random variables of the measurement outcomes, where we need to take into account that, e.g., child $\alpha$ cannot  simultaneously measure the POVMs corresponding to  $0_{\alpha}$ and $1_{\alpha}$. These unobservable covariances correspond to the black squares, while all white squares correspond to measurable covariances. Any scenario as in a) results in observable covariances that can be \emph{completed} into a matrix that can be decomposed into positive semidefinite blocks, analogous to Fig.~\ref{fig:triangle}. 
}
\end{figure}

Next we consider a generalization where each child $c_m$ can choose measurement settings $s_m$, modeled via a collection of POVMs $A^{(m,s_m)} = \{A^{(m,s_m)}_{x}\}_{x}$ (see Fig.~\ref{fig:triangleWithChoice}). The distribution of the measurement  outcomes $x_1,\ldots,x_M$, conditioned on the choices of  measurement settings $s_1,\ldots,s_M$, is
\begin{equation}
\label{QuantumWithInputs}
\begin{split}
&P(x_1,\ldots,x_M|s_1,\ldots,s_M)\\
& = \Tr([A^{(1,s_1)}_{x_1}\otimes\cdots\otimes A^{(M,s_M)}_{x_M}][\rho_{p_1}\otimes\cdots\otimes\rho_{p_N}]).
\end{split}
\end{equation}
We moreover associate a feature map  $Y^{(m,s_m)}$, to each child $c_m$ and each choice of measurement $s_m$, which maps into subspaces $\mathcal{V}_{m,s_m}$, each associated with a projector $P_{m,s_m}$.
In general, the different choices of POVMs correspond to non-commuting observables, thus implying that covariances of the form $\mathrm{Cov}(Y^{(m,s_m)},Y^{(m,s^{\prime}_{m})})$, for $s_{m}\neq s^{\prime}_{m}$,
are not empirically observable. In other words, these covariances cannot be determined experimentally, since child $c_m$ cannot simultaneously measure two non-commuting observables (see \cite{SuppMat} for further remarks). 
Based on the observable covariances only, we define the \emph{observable covariance matrix}
\begin{equation}
\label{dfbsfssbf}
\begin{split}
C_{\mathrm{observable}} := &   \sum_{m}\sum_{s_m}\mathrm{Cov}(Y^{(m,s_m)}) \\
 + & \sum_{m,m^{\prime}:m\neq m^{\prime}}\sum_{s_m,s^{\prime}_{m^{\prime}}} \mathrm{Cov}(Y^{(m,s_m)},Y^{(m^{\prime},s^{\prime}_{m^{\prime}})}),
\end{split}
\end{equation}
where the unobservable covariances correspond zero `blocks' $P_{m,s_m}C_{\mathrm{observable}}P_{m,s^{\prime}_{m}} = 0$ for $s_m\neq s^{\prime}_{m}$.
It turns out that one can complete the zero blocks in $C_{\mathrm{observable}}$ with a matrix $C_{\mathrm{completion}}$, such that $C := C_{\mathrm{observable}} + C_{\mathrm{completion}}$ is positive semidefinite. Moreover,  $C_{\mathrm{completion}}$ is non-zero \emph{only} on the blocks associated to the unobservable covariances, and thus 
\begin{equation}
\label{gfnsfgnsf}
\begin{split}
& P_{m,s_m}C_{\mathrm{completion}}P_{m,s_m} = 0,\\
& P_{m,s_m}C_{\mathrm{completion}}P_{m',s'_{m'}} = 0,\quad m\neq m'.
\end{split}
\end{equation}
The completed covariance matrix can be decomposed according to the DAG $G$, as detailed by the following proposition.
\begin{Proposition}
\label{PropQuantumBipartiteWithChoice}
Let the conditional distribution $P(x_1,\ldots,x_M|s_1,\ldots,s_M)$ be compatible, in the sense of (\ref{QuantumWithInputs}), with  the quantum causal structure $G$ with parents $p_1,\ldots,p_N$ and children $c_1,\ldots,c_M$, with associated inputs $s_1,\ldots,s_M$. Assume that each child $c_m$, and each input $s_m$, is assigned a feature map $Y^{(m,s_m)}$ into a vector space $\mathcal{V}_{m,s_m}$. Let the operator $C_{\mathrm{observable}}$  on $\mathcal{V} := \bigoplus_{m=1}^{M}\bigoplus_{s_m}\mathcal{V}_{m,s_m}$ be as defined in (\ref{dfbsfssbf}).
Then there exist operators  $C_{\mathrm{completion}}$, $R$ and $(C_n)_{n=1}^{N}$ on $\mathcal{V}$, such that $C_{\mathrm{completion}}$ satisfies  (\ref{gfnsfgnsf}) and
\begin{equation}
\label{ConditionCompletion}
C: = C_{\mathrm{observable}} + C_{\mathrm{completion}}\geq 0,
\end{equation}
and
\begin{eqnarray}
C & = & R + \sum_{n=1}^{N}C_n,\quad R\geq 0,\quad C_n\geq 0,\\
& & P^{(n)}C_n P^{(n)} = C_n,\quad R = \sum_{m,s_m}P_{m,s_m}R P_{m,s_m},\\
& & P^{(n)}:= \sum_{m\in\mathcal{C}_n}\sum_{s_m}P_{m,s_m},
\end{eqnarray}
where $P_{m,s_m}$ is the projector onto $\mathcal{V}_{m,s_m}$, and where $\mathcal{C}_n$ denotes the children of parent $p_n$ in the given DAG $G$.
\end{Proposition}
The proof for Proposition \ref{PropQuantumBipartiteWithChoice} and, for the sake of completeness, its classical counterpart, are provided in  \cite{SuppMat}.
The perhaps surprising conclusion is that even though we add inputs, the semidefinite test still only depends on the network topology, without distinguishing the classical and quantum case. As an application of Proposition \ref{PropQuantumBipartiteWithChoice}, we tested the nonlocal correlations arising from $\sigma_x$ and $\sigma_z$ measurements on a W-state \cite{Dur2000,Cabello2002,Chaves2011} $\ket{W}=(\sqrt{1/3})\left(\ket{001}+\ket{010}+\ket{100} \right)$ with visibility $v$, that is, $\rho_W=v\ket{W}\bra{W}+(1-v)\openone/8$. We observe numerically that these nonlocal correlations violate our SDP test for visibilities above $v \approx 3/4$, thus witnessing their incompatibility with the quantum triangle with inputs. In \cite{SuppMat} we moreover compare  the semidefinite test induced by Proposition \ref{PropQuantumBipartiteWithChoice}, with alternative tests with inputs, obtained via Proposition \ref{Prop:DecomposingQuantumNoInputs}.

In summary, we have presented a systematic method to test whether some observed covariances are compatible with a given quantum network topology. The approach is fairly general as it can be applied to any network where the correlations between the observed variables are mediated via independent sources, the number of inputs and outputs as well as the number of sources being arbitrary. Irrespective of Hilbert space dimensions and the type of quantum measurements being performed, our results show that the topology of the quantum network alone implies constraints on the covariance matrix of distributions compatible with it. Our method can be efficiently implemented via an SDP, even though the original problem is non-convex. Furthermore, it allows for analytical derivations of experimentally testable constraints that can be understood as device-independent witnesses of the topology of the quantum network. In comparison with another recently proposed test \cite{Renou2019}, our approach not only provides a significantly better description (see Figs. \ref{fig:triangle} and \ref{fig:4partite}) but can also be applied on a wider range of quantum networks. 

Given the ubiquitous role of quantum networks, we believe that our approach, together with other recently proposed alternatives \cite{Chaves2015a,wolfe2019quantum,Renou2019,navascues2020genuine,kraft2020quantum}, offer a novel suite of tools for network related problems, such as multipartite secure communication \cite{Lee2018}, distributed computing \cite{buhrman2010nonlocality}, quantum-repeaters \cite{briegel1998quantum}, or any other tasks where quantum networks might play a role.

An interesting open problem is whether our approach, based on the covariance of the observed correlations and that provides an outer approximation to the true set of correlations compatible with a given network, can be generalized to include higher-order moments of the distribution, thus providing a tighter description of the quantum set of correlations. Also, one can wonder whether the same constraints hold for non-signalling correlations \cite{gisin2020constraints} and if our method can be combined with inflation techniques \cite{wolfe2019quantum}. We hope that our results will trigger further developments in these directions.

\begin{acknowledgments}
We thank an anonymous referee for pointing out the possibility to use Proposition \ref{Prop:DecomposingQuantumNoInputs} to   obtain alternative tests with inputs.  
J{\AA} is supported by the Excellence Initiative of the German Federal and State Governments (Grant ZUK 81), the ARO under contract W911NF-14-1-0098 (Quantum Characterization, Verification, and Validation), and the DFG (SPP1798 CoSIP).
Funded by the Deutsche Forschungsgemeinschaft (DFG, German Research Foundation) under Germany's
Excellence Strategy - Cluster of Excellence Matter and Light for Quantum Computing (ML4Q) EXC 2004/1-390534769.
RC and RN  acknowledge the John Templeton Foundation via the Grant Q-CAUSAL No. 61084, the Serrapilheira Institute (Grant No. Serra-1708-15763), the Brazilian National Council for Scientific and Technological Development (CNPq) via the National Institute for Science and Technology on Quantum Information (INCT-IQ) and Grants No. 307172/2017-1 and No. 406574/2018-9, the Brazilian agencies MCTIC and MEC. CD was supported by a fellowship from the Grand Challenges Initiative at Chapman University. This project/research was supported by grant number FQXi-RFP-IPW-1905 from the Foundational Questions Institute and Fetzer Franklin Fund, a donor advised fund of Silicon Valley Community Foundation
\end{acknowledgments}

\bibliography{biblio}

\end{document}

\title{Supplemental Material for ``Semidefinite tests for quantum network topologies"}

\author{Johan {\AA}berg}
\affiliation{Institute for Theoretical Physics, University of Cologne, Germany}

\author{Ranieri Nery}
\affiliation{International Institute of Physics, Federal University of Rio Grande do Norte, 59070-405 Natal, Brazil}

\author{Cristhiano Duarte}
\affiliation{
 Schmid College of Science and Technology, Chapman University 
  1 University Dr, Orange, CA 92866
}

\author{Rafael Chaves}
\affiliation{International Institute of Physics, Federal University of Rio Grande do Norte, 59070-405 Natal, Brazil}
\affiliation{School of Science and Technology, Federal University of Rio Grande do Norte, 59078-970 Natal, Brazil}

\maketitle

\onecolumngrid

\section{\label{SecClassicalWithoutInputs} Brief summary: Semidefinite decompositions for classical networks}
For convenience, we here restate the core result on the classical semidefinite test obtained in \cite{KelaEtAl20}. As explained in the main text,  we exclusively consider the class of DAGs where the vertices can be partitioned into a set of parents that is distinct from the set of children,
and where the arrows only go from parent vertices to children vertices. This class of DAGs is in \cite{KelaEtAl20} referred to as `bipartite'. However, here we avoid this terminology, since in the quantum context, this may be misinterpreted as each parent having at most two children, i.e., that each parent only can create bipartite entanglement, while we here allow for scenarios where the parents can create quantum correlations between any number of children.

For this class of DAGs with parents $p_1,\ldots,p_N$ and children $c_1,\ldots,c_M$, the joint distribution of the parents (or latent variables) and children (observables) can be written
\begin{equation}
\label{ClassicalWithoutInputs}
P(c_1,\ldots,c_M,p_1,\ldots,p_N) =  \Pi_{m=1}^{M}P(c_m|\mathcal{P}_m)\Pi_{n=1}^{N}P(p_n),
\end{equation}
where $\mathcal{P}_m$ denotes the set of parents of child $c_m$. 
We say that an observed distribution $P(c_1,\ldots,c_M)$ is compatible with the given classical causal structure $G$ with parents $p_1,\ldots,p_N$ and children $c_1,\ldots, c_M$ (or that $G$ is a causal explanation of the distribution), if it is the margin of some distribution of the form (\ref{ClassicalWithoutInputs}). In \cite{KelaEtAl20}, it was shown that this structure induces a particular signature on any covariance matrix that can be construed from the random observables $c_1,\ldots,c_M$. We form covariance matrices by assigning feature maps $Y^{(m)} := Y^{(m)}(c_m)$ into pairwise orthogonal vector spaces $\mathcal{V}_m$. On the joint space $\mathcal{V} :=\bigoplus_{m=1}^{M}\mathcal{V}_m$, we let $P_m$ denote the projector onto $\mathcal{V}_m$.

The following proposition  is taken from \cite{KelaEtAl20} (rephrased in order to better fit our discussions).
\begin{Proposition}[\cite{KelaEtAl20}]
\label{PropClassicalWithoutInputs}
Let the distribution $P(c_1,\ldots,c_M)$ be compatible with the classical causal structure $G$ with parents $p_1,\ldots,p_N$ and children $c_1,\ldots,c_M$. Assume that for each child $c_m$ there is assigned a feature map $Y^{(m)}$ into a vector space $\mathcal{V}_m$. 
Then there exist operators $R$ and $(C_n)_{n=1}^{N}$ on $\mathcal{V} := \bigoplus_{m=1}^{M}\mathcal{V}_{m}$, such that
 the covariance matrix of $Y= \sum_{m=1}^{M}Y^{(m)}$ satisfies
\begin{equation}
\label{DecCovClassical}
\mathrm{Cov}(Y) = R + \sum_{n=1}^{N}C_n,\quad R\geq 0,\quad C_n\geq 0,
\end{equation}
where
\begin{equation}
\label{ConProjClassical}
P^{(n)}C_nP^{(n)} = C_n,\quad R =\sum_{m=1}^{M}P_mRP_m,
\end{equation}
for the projectors $P^{(n)} := \sum_{m\in \mathcal{C}_n}P_m$, where $\mathcal{C}_n$ denotes the children of parent $p_n$ in the given DAG $G$, and where $P_m$ is the projector onto $\mathcal{V}_m$.

\end{Proposition}

\emph{Remark:} From this proposition we can conclude that whenever we find a covariance matrix that cannot be decomposed in this manner, then we can exclude $G$ as an explanation of the observed covariance or underlying distribution. This necessary condition can be used to design an SDP-test that must be satisfied whenever $G$ is a causal explanation, as discussed in \cite{KelaEtAl20} (see also the discussion about the quantum counterpart in Section \ref{SecWitness}).

\emph{Remarks on feature maps:} As mentioned in the main text, the mappings from the measurement outcomes $x_m$ to the vectors $Y^{(m)}_{x_m}$ are sometimes referred to as feature maps \cite{Schoelkopf2002learning}.  Intuitively, it seems reasonable that it would be enough to map distinct measurement outcomes to an orthonormal basis. This intuition was confirmed in \cite{KelaEtAl20}, where it was shown that for a finite number of  outcomes, it suffices to associate these to linearly independent vectors (referred to as  `universal' feature maps in \cite{KelaEtAl20}). For finite number of outcomes, it is thus enough to choose orthonormal feature-vectors, which also is what has been done throughout this paper for all concrete examples. The freedom to  choose feature maps can nevertheless provide a useful flexibility. If one considers a countably infinite number of outcomes, or more generally a continuum of outcomes,  one may potentially construct generalizations of the notion of universal feature maps. Although this may yield a semidefinite test that could be useful from a theoretical point of view, it seems challenging to obtain practically functioning numerical implementations. It is also not difficult to imagine scenarios where the number of outcomes are finite, but too many for  performing the resulting tests in practice. However, since the semidefinite test is valid for any choice of feature maps, we can still obtain finite-dimensional covariance matrices, and practically feasible tests, by mapping distinct outcomes to  linearly dependent feature-vectors. In other words, we sacrifice potential power of the test in order to gain implementability.

%%%%%%%%%%%%%%%%%%%%%%%%%%%%%%%%%%%%%%%%%%%%%%%%%%%%%%%%%%

\section{\label{SecProofProp1}Proof of  Proposition 1 in the main text}

Before we turn to the setting and proof of Proposition 1 in the main text, we first establish a technical lemma that not only will be important for the proof of Proposition 1, but also will play a similar role in Section \ref{SecProofOfBipartWithChoice}, when we prove Proposition 2 of the main text.

\subsection{Technical Lemma} \label{SubSec:TechnicalLemma}

For a Hilbert space $\mathcal{H}$, we let $\mathcal{S}(\mathcal{H})$ denote the set of density operators on $\mathcal{H}$. 
Given a family $\mathcal{H}^{p_1},\ldots,\mathcal{H}^{p_N}$ of Hilbert spaces, let  
\begin{equation}
\label{fdbsfb}
\mathcal{H} :=\bigotimes_{n=1}^{N}\mathcal{H}^{p_n}.
\end{equation}
In the following we consider linear operators $Q:\mathcal{H}\rightarrow \mathcal{H}\otimes\mathcal{V}$. For density operators $\rho$ on $\mathcal{H}$, we often consider objects such as $\Tr_{\mathcal{H}}(Q\rho)$, $\Tr_{\mathcal{H}}(Q^{\dagger}\rho)$  and $\Tr_{\mathcal{H}}(QQ^{\dagger}\rho)$, where $\Tr_{\mathcal{H}}$ denotes the partial trace with respect to $\mathcal{H}$. While $\Tr_{\mathcal{H}}(Q\rho)$ is an element of $\mathcal{V}$, and  $\Tr_{\mathcal{H}}(Q^{\dagger}\rho)$ is an element of the dual $\mathcal{V}^{\ast}$,  $\Tr_{\mathcal{H}}(QQ^{\dagger}\rho)$ is  an element of the space $\mathcal{L}(\mathcal{V})$  of linear operators  on $\mathcal{V}$. More specifically, if one chooses 
$\{\ket{\psi_{k}}\}_{k}$ as an orthonormal basis for $\mathcal{H}$ and $\{\ket{a_{j}}\}_{j}$ as a basis for $\mathcal{V}$, then any linear operator $Q:\mathcal{H} \rightarrow \mathcal{H} \otimes \mathcal{V}$ decomposes as
\begin{align}
Q=\sum_{j}\sum_{k,l}Q^{j}_{k,l} \ket{\psi_{k}}\ket{a_j}\bra{\psi_{l}}.
\end{align}
The partial traces thus become
\begin{equation}
\begin{split}
\Tr_{\mathcal{H}}(Q\rho)  = & \sum_j \sum_{k,l}  Q^{j}_{k,l} \langle\psi_l|\rho|\psi_k\rangle|a_j\rangle\in\mathcal{V},\\
\Tr_{\mathcal{H}}(Q^{\dagger}\rho)  = & \sum_j \sum_{k,l} (Q^{j}_{k,l})^{*}\langle\psi_k|\rho|\psi_l\rangle\langle a_j|\in\mathcal{V}^{\ast},\\
\Tr_{\mathcal{H}}(QQ^{\dagger}\rho) = & \sum_{l} \sum_{j,j'} \sum_{k,k'} Q^{j}_{k,l}  (Q^{j'}_{k',l})^{*}  \langle\psi_{k'}|\rho|\psi_k\rangle |a_j\rangle\langle a_{j'}|\in\mathcal{L}(\mathcal{V}).
\end{split}
\end{equation}

\begin{Lemma}
\label{nfkelnlfk}
Let $Q:\mathcal{H}\rightarrow \mathcal{H}\otimes\mathcal{V}$ be a  linear operator, and let $\rho_n\in\mathcal{S}(\mathcal{H}^{p_n})$ for each $n=1,\ldots, N$. Define $\rho := \rho_{p_1}\otimes\cdots\otimes\rho_{p_N}$. Then
\begin{equation}
\begin{split}
\Tr_{\mathcal{H}}(QQ^{\dagger}\rho)- \Tr_{\mathcal{H}}(Q\rho)\Tr_{\mathcal{H}}(Q^{\dagger}\rho) = \sum_{n=1}^{N}C_{n},
\end{split}
\label{Eq.FirstSum}
\end{equation}
where
\begin{equation}
\label{nfdlkbnfd}
\begin{split}
 C_1 := & \Tr_{p_1,\ldots,p_N}\bigg[\Big(Q- \hat{1}_{p_1}\otimes\Tr_{p_1}(Q\rho_{p_1}) \Big)\Big(Q- \hat{1}_{p_1}\otimes\Tr_{p_1}(Q\rho_{p_1}) \Big)^{\dagger}[\rho_{p_1}\otimes\cdots\otimes \rho_{p_N}]\bigg],\\
& \\
 C_n := & \Tr_{p_n,\ldots,p_N}\bigg[\bigg(\Tr_{p_1,\ldots, p_{n-1}}(Q[\rho_{p_1}\otimes\cdots \otimes\rho_{p_{n-1}}]) - \hat{1}_{p_n}\otimes\Tr_{p_1,\ldots,p_n}(Q[\rho_{p_1}\otimes\cdots \otimes\rho_{p_n}]) \bigg)\\
& \quad\quad\quad\quad\quad\bigg(\Tr_{p_1,\ldots,p_{n-1}}(Q[\rho_{p_1}\otimes\cdots \otimes \rho_{p_{n-1}}]) - \hat{1}_{p_n}\otimes\Tr_{p_1,\ldots,p_n}(Q[\rho_{p_1}\otimes\cdots \otimes \rho_{p_n}]) \bigg)^{\dagger}\\
& \quad\quad\quad\quad\quad\quad  [\rho_{p_n}\otimes\cdots\otimes \rho_{p_N}]\bigg], \quad\quad\quad 2\leq n\leq N-1,\\
& \\
C_{N} :=&  \Tr_{p_N}\bigg[\bigg(\Tr_{p_1,\ldots, p_{N-1}}(Q[\rho_{p_1}\otimes\cdots \otimes\rho_{p_{N-1}}]) - \hat{1}_{p_N}\otimes\Tr_{p_1,\ldots,p_N}(Q[\rho_{p_1}\otimes\cdots \otimes\rho_{p_N}]) \bigg)\\
& \quad\quad\quad \bigg(\Tr_{p_1,\ldots,p_{N-1}}(Q[\rho_{p_1}\otimes\cdots \otimes \rho_{p_{N-1}}]) - \hat{1}_{p_N}\otimes\Tr_{p_1,\ldots,p_N}(Q[\rho_{p_1}\otimes\cdots \otimes \rho_{p_N}]) \bigg)^{\dagger}\rho_{p_N}\bigg],\\
\end{split}
\end{equation}
One can realize that each $C_n$ is a positive semi-definite operator on $\mathcal{V}$.
\end{Lemma}
\begin{proof}
We break this proof into two main pieces. First of all, one should note that Eq.~\eqref{Eq.FirstSum} can be rewritten as the telescopic sum  
\begin{equation}
\begin{split}
\Tr(QQ^{\dagger}\rho)- \Tr(Q\rho)\Tr(Q^{\dagger}\rho) = \sum_{n=1}^{N}D_{n},
\end{split}
\end{equation}
where
\begin{equation}
\begin{split}
D_{1} := &
\Tr_{p_1,\ldots,p_N}\big[QQ^{\dagger}[\rho_{p_1}\otimes\cdots\otimes \rho_{p_N}]\big] 
-\Tr_{p_2,\ldots,p_N}\big[\Tr_{p_1}(Q\rho_{p_1})  \Tr_{p_1}(Q^{\dagger}\rho_{p_1}) [\rho_{p_2}\otimes\cdots\otimes \rho_{p_N}]\big],\\
D_{n} := & 
\Tr_{p_n,\ldots,p_N}\bigg[\Tr_{p_1,\ldots, p_{n-1}}(Q[\rho_{p_1}\otimes\cdots \otimes\rho_{p_{n-1}}]) \Tr_{p_1,\ldots,p_{n-1}}(Q^{\dagger}[\rho_{p_1}\otimes\cdots \otimes \rho_{p_{n-1}}])[\rho_{p_n}\otimes\cdots\otimes \rho_{p_N}]\bigg]\\
& 
-\Tr_{p_{n+1},\ldots,p_N}\bigg[\Tr_{p_1,\ldots, p_n}(Q[\rho_{p_1}\otimes\cdots \otimes\rho_{p_n}])  \Tr_{p_1,\ldots,p_n}(Q^{\dagger}[\rho_{p_1}\otimes\cdots \otimes \rho_{p_n}]) [\rho_{p_{n+1}}\otimes\cdots\otimes \rho_{p_N}]\bigg],\\ 
& \quad\quad\quad 2\leq n\leq N-1,\\
D_{N} := & \Tr_{p_N}\bigg[\Tr_{p_1,\ldots, p_{N-1}}(Q[\rho_{p_1}\otimes\cdots \otimes\rho_{p_{N-1}}]) \Tr_{p_1,\ldots,p_{N-1}}(Q^{\dagger}[\rho_{p_1}\otimes\cdots \otimes \rho_{p_{N-1}}])\rho_{p_N}\bigg]\\
& 
-\Tr_{p_1,\ldots, p_N}\big[Q[\rho_{p_1}\otimes\cdots \otimes\rho_{p_N}]\big]  \Tr_{p_1,\ldots,p_N}\big[Q^{\dagger}[\rho_{p_1}\otimes\cdots \otimes \rho_{p_N}]\big].
\end{split}
\end{equation}
The second step consists in noticing that for each $n  = 1,\ldots,N$ one has $D_n = C_n$ with $C_n$ as in (\ref{nfdlkbnfd}).
\end{proof}

\subsection{Semidefinite decompositions for quantum networks}

Let $G$ be a  graph with $N$ parents $p_1,\ldots,p_N$, and $M$ children $c_1,\ldots,c_M$. Recall that $\mathcal{C}_n$ denotes the set of children with parent $p_n$, and $\mathcal{P}_m$ denotes the set of parents to child $c_m$. Also recall that we here restrict to the class of graphs $G$ where there only are arrows directed from the set of parents to the set of children, i.e., no parent is itself a child, and no child itself a parent.

For each parent $p_n$ we assume a collection of Hilbert spaces $\{\mathcal{H}_{m}^{n}\}_{m\in\mathcal{C}_n}$, and define
\begin{equation}
\label{Structure}
\begin{split}
\mathcal{H}_{c_m} := &\bigotimes_{n\in\mathcal{P}_m}\mathcal{H}^{n}_m,\quad\mathcal{H}^{p_n} := \bigotimes_{m\in\mathcal{C}_n}\mathcal{H}^{n}_m,\\
\mathcal{H} := &  \bigotimes_{m=1}^{M}\bigotimes_{n\in\mathcal{P}_m}\mathcal{H}^n_m = \bigotimes_{n=1}^{N}\bigotimes_{m\in\mathcal{C}_n}\mathcal{H}^{n}_m = \bigotimes_{m=1}^{M}\mathcal{H}_{c_m} = \bigotimes_{n=1}^{N}\mathcal{H}^{p_n}.
\end{split}
\end{equation}
The interpretation of this structure is that $\mathcal{H}^{n}_{m}$ is the Hilbert space of the system that parent $p_n$ sends to child $c_m$ (if $p_n$ indeed is a parent of $c_m$).

We let $\rho_{n}\in\mathcal{S}(\mathcal{H}^{p_n})$ for $n = 1,\ldots,N$, and the define the joint state 
\begin{equation}
\label{DAGGstates}
\rho:= \rho_{p_1}\otimes\cdots\otimes\rho_{p_N}\in\mathcal{S}(\mathcal{H}).
\end{equation} 
Hence, the parents are all uncorrelated.

On each of the Hilbert spaces $\mathcal{H}_{c_m}$ we assign a POVM $\{A^{(m)}_{x_m}\}_{x_m}$. 
From the collection of POVMs $\{A^{(1)}_{x_1}\}_{x_1},\ldots, \{A^{(M)}_{x_M}\}_{x_M}$, we define the  joint POVM
 \begin{equation}
 \label{DAGGPOVMs}
A_{x_1,\ldots,x_M} := A^{(1)}_{x_1}\otimes\cdots\otimes A^{(M)}_{x_M}.
\end{equation}
By measuring this joint POVM on the state $\rho$, it follows that the outcome (or observation)  $x_1,\ldots,x_M$ occurs with the probability 
\begin{equation}
\label{QuantumWithoutInputsAgain}
P(x_1,\ldots,x_M) = \Tr([A^{(1)}_{x_1}\otimes\cdots\otimes A^{(M)}_{x_M}][\rho_{p_1}\otimes\cdots\otimes\rho_{p_N}]).
\end{equation}
We say that an observed distribution $P(x_1,\ldots,x_M)$ is compatible with the given quantum casual structure $G$ with parents $p_1,\ldots,p_N$ and children $c_1,\ldots,c_M$, if it can be written as in (\ref{QuantumWithoutInputsAgain}), for some state as in  (\ref{DAGGstates}) and some POVM as in  (\ref{DAGGPOVMs}).

Analogous to the classical case in Section \ref{SecClassicalWithoutInputs}, we assign a feature map $Y^{(m)}$ to each child $c_m$, i.e., to each measurement outcome $x_m$, corresponding to POVM element $A^{(m)}_{x_m}$, we associate an element $Y^{(m)}_{x_m}$ in a vector space $\mathcal{V}_m$. We also define $\mathcal{V} := \bigoplus_{m=1}^{M}\mathcal{V}_m$. 
 The joint feature map for the joint measurement of all children is
\begin{equation*}
Y_{x_1,\ldots,x_M}:= Y^{(1)}_{x_1}+\cdots +Y^{(M)}_{x_M} \in\mathcal{V}.
\end{equation*}
 We let $P_{m}$ denote the projector onto the subspace $\mathcal{V}_m$, and for each $n = 1,\ldots, N$ we define the projector
\begin{equation}
\label{sfbsfgb}
P^{(n)} := \sum_{m\in\mathcal{C}_n}P_{m}.
\end{equation}

\begin{figure}
\begin{center}
 \includegraphics[width=0.9\columnwidth]{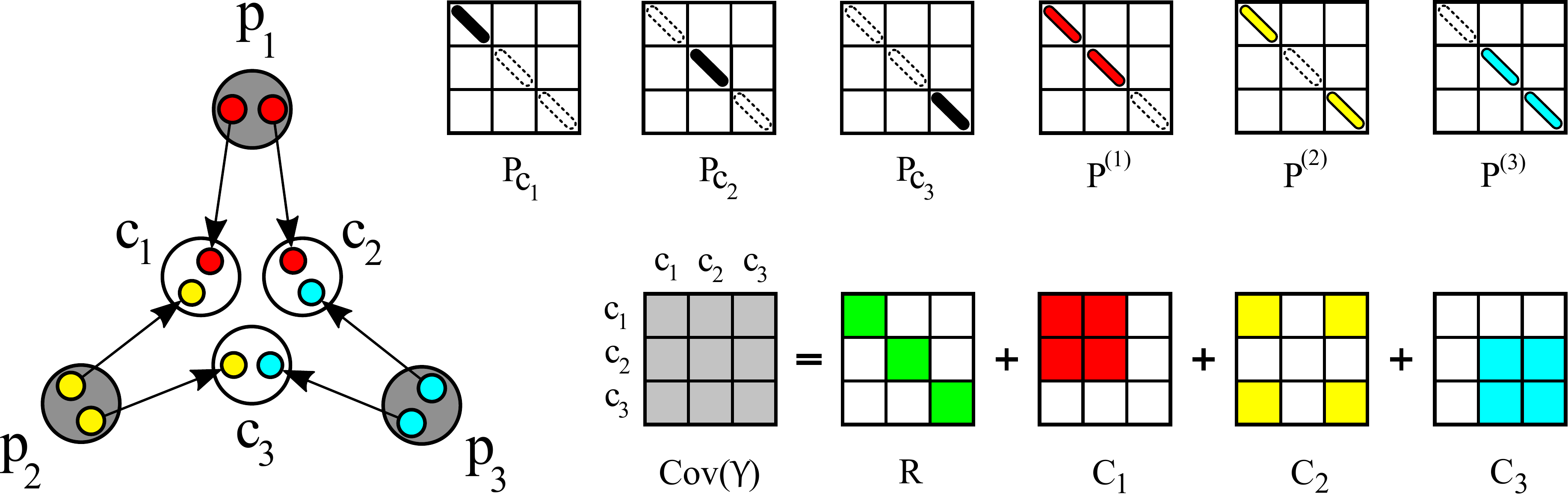}
\end{center} 
\caption{\label{fig:BlockStructure} 
{\bf Block structure in the semidefinite decomposition for the triangular DAG.} 
To each child $c_1$, $c_2$, $c_3$ we associate a subspace $\mathcal{V}_1$, $\mathcal{V}_2$, $\mathcal{V}_3$ with a corresponding projector $P_{c_1}$, $P_{c_2}$, $P_{c_3}$.
To parent $p_1$ we associate a projector onto the subspaces of its children, $P^{(1)} :=  P_{c_1} +P_{c_2}$,  for parent $p_2$  the projector  $P^{(2)} := P_{c_1} +P_{c_3}$, and for parent $p_3$ the projector $P^{(3)} := P_{c_2} + P_{c_3}$. One can view the semidefinite decomposition $\mathrm{Cov}(Y) = R + C_1 + C_2 +C_3$ in terms of a block-structure with respect to the subspaces $\mathcal{V}_{1}$, $\mathcal{V}_{2}$, $\mathcal{V}_{3}$. The operator $R$ is block-diagonal, in the sense that $R= P_{c_1}RP_{c_1} + P_{c_2}RP_{c_2} + P_{c_3}RP_{c_3}$.  The operator $C_1$ only has support on the subspace $\mathcal{V}_1\oplus\mathcal{V}_2$, i.e., $P^{(1)}C_1 P^{(1)} = C_2$, and analogously $P^{(2)}C_2 P^{(2)} = C_2$, and $P^{(3)}C_3 P^{(3)} = C_3$.
}
\end{figure}

For the sake of convenience, we here restate Proposition 1 as in the main text. We  emphasize that the semidefinite decomposition  (\ref{vdfbvsdfagain}) can be understood via a block-structure of $\mathrm{Cov}(Y)$, with respect to the subspaces $\mathcal{V}_m$. This is illustrated for the triangular DAG in Fig.~\ref{fig:BlockStructure}.

\begin{Proposition}
\label{Prop:DecomposingQuantumNoInputsAgain}
Let the distribution $P(x_1,\ldots,x_M)$ be compatible, in the sense of  (\ref{QuantumWithoutInputsAgain}), with a quantum causal structure $G$ with parents $p_1,\ldots,p_N$ and children $c_1,\ldots,c_M$.
 Assume that each child $c_m$  is assigned a feature map $Y^{(m)}$ into a vector space $\mathcal{V}_m$. 
Then there exist operators $R$ and $(C_n)_{n=1}^{N}$ on $\mathcal{V} := \bigoplus_{m=1}^{M}\mathcal{V}_{m}$, such that
 the covariance matrix of $Y= \sum_{m=1}^{M}Y^{(m)}$ satisfies
\begin{eqnarray}
\label{vdfbvsdfagain}
\mathrm{Cov}(Y) & = &  R + \sum_{n=1}^{N}C_n,\quad R\geq 0,\quad C_n\geq 0,\\
\label{Eq.ThmProjectorsCAndRagain}
& & P^{(n)}C_nP^{(n)} =C_n,\quad R = \sum_{m=1}^{M}P_mRP_m,
\end{eqnarray}
for the projectors $P^{(n)} := \sum_{m\in \mathcal{C}_n}P_m$, where $\mathcal{C}_n$ denotes the children having parent $p_n$ in the given DAG $G$, and where $P_m$ is the projector onto $\mathcal{V}_m$.
\end{Proposition}

\emph{Remark concerning $R$:} Note that since $R$ is block-diagonal with respect to the projectors $P_m$, we can in many cases incorporate the blocks $P_mRP_m$ into the operators $C_n$.  In other words, we create a new set of operators $C'_n$ that satisfy all the requirements, and let $R = 0$. This is possible whenever the collection of supports of the projectors $P^{(n)}$ spans the entire space $\mathcal{V}$, where the exception is when there exists a child $c_m$ that does not have any parent. (This observation also applies to the classical setting, as was briefly remarked in \cite{KelaEtAl20}.) Hence, except in the special case of orphans (i.e. children nodes with no parents in the DAG), it is sufficient for the semidefinite test to  search among $C_n$ that satisfy all conditions of Proposition \ref{Prop:DecomposingQuantumNoInputsAgain}, while putting $R = 0$.

\begin{proof}
By the assumption that $P(x_1,\ldots,x_M)$ be compatible with the quantum causal structure $G$ with parents $p_1,\ldots,p_N$ and children $c_1,\ldots,c_M$,  there exists a state on the form (\ref{DAGGstates}), and a POVM of the form (\ref{DAGGPOVMs}), such that $P(x_1,\ldots,x_M)$ can be written as in (\ref{QuantumWithoutInputsAgain}). With the assigned feature maps $Y^{(m)}$ each outcome $x_m$ of the measurement of the POVM $\{A^{(m)}_{x_m}\}_{x_m}$ corresponds to a vector in $Y^{(m)}_{x_m}$. The measurement of the POVM $A_{x_1,\ldots,x_M}$ thus results in a random vector $Y$ on $\mathcal{V}$.
In the following, we will show that the covariance matrix $\mathrm{Cov}(Y)$ can be decomposed as in (\ref{vdfbvsdfagain})  for $R\geq 0$ and $C_n\geq 0$.
Define $Q\in \mathcal{V}\otimes\mathcal{L}(\mathcal{H})$ by
\begin{equation}
\label{bsfgbsfg}
\begin{split}
Q := & \sum_{x_1,\ldots,x_M} Y_{x_1,\ldots, x_M}\otimes A^{(1)}_{x_1}\otimes\cdots\otimes A^{(M)}_{x_M} \\
= & \sum_{x_1,\ldots,x_M} ( Y^{(1)}_{x_1}+\cdots +Y^{(M)}_{x_M})\otimes A^{(1)}_{x_1}\otimes\cdots\otimes A^{(M)}_{x_M} \\
= & \sum_{m=1}^{M} \sum_{x_m} Y^{(m)}_{x_m}\otimes A^{(m)}_{x_m}\otimes \hat{1}_{/c_m},
\end{split}
\end{equation}
where $\hat{1}_{/c_m}$ denotes the identity operator on $\mathcal{H}_{c_1}\otimes\cdots\otimes \mathcal{H}_{c_{m-1}}\otimes \mathcal{H}_{c_{m+1}}\otimes\cdots\otimes\mathcal{H}_{c_M}$, and where the last equality follows from $\{ A^{(m)}_{x_m}\}_{x_m}$ being a POVM. Next we note that
\begin{equation}
\begin{split}
\textrm{Cov}(Y) = & E(YY^{\dagger})-E(Y)E(Y)^{\dagger}\\
= & \sum_{x_1,\ldots,x_M}  Y_{x_1,\ldots,x_M}Y^{\dagger}_{x_1,\ldots,x_M} \Tr(A_{x_1,\ldots,x_M}\rho)  \\
& - \sum_{x_1,\ldots,x_M} Y_{x_1,\ldots,x_M} \Tr(A_{x_1,\ldots,x_M}\rho)   \sum_{x'_1,\ldots,x'_M}  Y_{x'_1,\ldots,x'_M}^{\dagger} \Tr(A_{x'_1,\ldots,x'_M}\rho) \\
= & \sum_{x_1,\ldots,x_M}  Y_{x_1,\ldots,x_M}Y^{\dagger}_{x_1,\ldots,x_M} \Tr(A_{x_1,\ldots,x_M}\rho) \\
& - \Tr_{\mathcal{H}}(Q\rho)\Tr_{\mathcal{H}}(Q^{\dagger}\rho)\\
=  & R +\Tr_{\mathcal{H}}(QQ^{\dagger}\rho) - \Tr_{\mathcal{H}}(Q\rho)\Tr_{\mathcal{H}}(Q^{\dagger}\rho)\\
=  &R +\sum_{n=1}^{N}C_n,
\end{split}
\end{equation}
where the last equality follows by Supplementary Lemma \ref{nfkelnlfk}, and where 
\begin{equation}
\label{dfbsfgb}
\begin{split}
R:= & \sum_{x_1,\ldots,x_M}  Y_{x_1,\ldots,x_M}Y^{\dagger}_{x_1,\ldots,x_M} \Tr(A_{x_1,\ldots,x_M}\rho)  - \Tr_{\mathcal{H}}(QQ^{\dagger}\rho),
\end{split}
\end{equation}
and where $C_n$ is as in Supplementary Lemma \ref{nfkelnlfk}, which also yields $C_n\geq 0$.

Next we shall show that  $P^{(n)}C_nP^{(n)} =C_n$, for the projectors $P^{(n)}$ defined with respect to the given DAG $G$.
For $2\leq n\leq N-1$, recall the definition of $C_n$ in (\ref{nfdlkbnfd}). With the definition of $Q$ in (\ref{bsfgbsfg}), it follows that 
\begin{equation}
\label{biouztotiz}
\begin{split}
C_n = &  \sum_{m,m'=1}^{M}\sum_{x_{m},x'_{m'}}  Y^{(m)}_{x_m} {Y^{(m')}_{x'_{m'}}}^{\dagger}   \Tr_{p_n,\ldots,p_N}\bigg[W^{m,n}_{x_m}{W^{m',n}_{x'_{m'}}}^{\dagger} [\rho_{p_n}\otimes\cdots\otimes \rho_{p_N}]\bigg],
\end{split}
\end{equation}
where 
\begin{equation}
\label{dsfbsfgb}
\begin{split}
W^{m,n}_{x_m} 
:= &\Tr_{p_1,\ldots, p_{n-1}}([A^{(m)}_{x_m}\otimes \hat{1}_{/c_m}][\rho_{p_1}\otimes\cdots \otimes\rho_{p_{n-1}}])\\
&  - \hat{1}_{p_n}\otimes\Tr_{p_1,\ldots,p_n}([A^{(m)}_{x_m}\otimes \hat{1}_{/c_m}][\rho_{p_1}\otimes\cdots \otimes\rho_{p_n}]).
\end{split}
\end{equation}
Suppose $m\notin \mathcal{C}_n$. This implies that $\hat{1}_{p_n}\otimes \Tr_{p_n}([A_{x_m}^{(m)}\otimes\hat{1}_{/c_m}]\rho_{p_n})= A_{x_m}^{(m)}\otimes\hat{1}_{/c_m}$.
Consequently $W^{m,n}_{x_m}  = 0$, for all $x_m$.
Hence, if $m\notin \mathcal{C}_n$ or if $m'\notin \mathcal{C}_n$, then
\begin{equation}
\label{nghndgh}
W^{m,n}_{x_m}{W^{m',n}_{x'_{m'}}}^{\dagger} = 0,\quad \forall x_m, x'_{m'}.
\end{equation}
Recall that $Y^{(m)}_{x_m}$ is supported on $\mathcal{V}_m$. By comparing (\ref{nghndgh}) with with (\ref{biouztotiz}) one can note that $P_m C_n P_{m'} = 0$ if $m\notin \mathcal{C}_n$ or if $m'\notin \mathcal{C}_n$. 
Thus 
\begin{equation}
\begin{split}
C_n = & \sum_{m,m'=1}^{M} P_m C_n P_{m'}\\
 = & \sum_{m,m'\in \mathcal{C}_n}^{M} P_m C_n P_{m'}\\
= & P^{(n)}C_n P^{(n)}.
\end{split}
\end{equation} 
We can conclude that $P^{(n)}C_nP^{(n)} = C_n$, for $2\leq n\leq N-1$. The proofs for the cases $n = 1$ and $n = N$ are analogous.

Next we show that $R$, as defined in Eq.~\eqref{dfbsfgb}, is positive semidefinite.
Recall that $R$ is an operator on $\mathcal{V}$. 
In addition, from the definition of $R$, the definition of $Q$ in \eqref{bsfgbsfg}, and the fact that $\{A_{x_1,\ldots, x_M}\}_{x_1,\ldots, x_M}$ and thus $\sum_{x_1,\ldots,x_M} A_{x_1,\ldots, x_M}  = \hat{1}$, we have 
\begin{equation}
\label{rmrimruimr}
\begin{split}
R = &\sum_{x_1,\ldots,x_M} \Tr_{\mathcal{H}}\bigg[\bigg( Y_{x_1,\ldots,x_M}\otimes \hat{1} -  \sum_{\overline{x}_1,\ldots,\overline{x}_M}  Y_{\overline{x}_1,\ldots,\overline{x}_M}\otimes
A_{\overline{x}_1,\ldots, \overline{x}_M}   \bigg)\\
&   A_{x_1,\ldots, x_M}  \bigg( Y_{x_1,\ldots,x_M}\otimes \hat{1} -   \sum_{\overline{x}'_1,\ldots \overline{x}'_M}  Y_{\overline{x}'_1,\ldots,\overline{x}'_M}
\otimes A_{\overline{x}'_1,\ldots, \overline{x}'_M}   \bigg)^{\dagger}\rho \bigg] 
\end{split}
\end{equation}
and hence, for arbitrary $c \in \mathcal{V}$ it is the case that 
\begin{equation}
\begin{split}
c^{\dagger}Rc  = & \sum_{x_1,\ldots,x_M} \Tr[Z_{x_1,\ldots,x_M}A_{x_1,\ldots, x_M} Z_{x_1,\ldots,x_M}^{\dagger}\rho],\\
Z_{x_1,\ldots,x_M} := &  c^{\dagger}Y_{x_1,\ldots,x_M}  \otimes \hat{1} -  \sum_{\overline{x}_1,\ldots,\overline{x}_M}  c^{\dagger}Y_{\overline{x}_1,\ldots,\overline{x}_M}
A_{\overline{x}_1,\ldots, \overline{x}_M}.
\end{split}
\end{equation}
Since $A_{x_1,\ldots, x_M}\geq 0$ and $\rho\geq 0$ it follows that $c^{\dagger}Rc \geq 0$, and thus $R\geq 0$.

The final step is to show that $R$ satisfies the decomposition $R = \sum_{m=1}^{M}P_mRP_m$, where $P_m$ is the projector onto $\mathcal{V}_m$ in $\bigoplus_{m=1}^{M}\mathcal{V}_m$.
Recall that $A_{x_1,\ldots,x_M} = A_{x_1}^{(1)}\otimes\cdots\otimes A^{(M)}_{x_M}$ and $Y_{x_1,\ldots,x_M} = Y^{(1)}_{x_1} +\cdots+Y^{(M)}_{x_M}$, where $Y^{(m)}_{x_m}$ has support in $\mathcal{V}_m$. By combining these observations with Eq.~\eqref{rmrimruimr} for $m\neq m'$ we get

\begin{equation}
\begin{split}
\label{Eq:PmRPmprime}
P_{m}RP_{m'} =& \sum_{x_1,..., x_M} \Tr\bigg[ \bigg( Y_{x_m}^{(m)} \otimes \hat{1} - \sum_{\overline{x}_m} Y_{\overline{x}_m}^{(m)} \otimes A_{\overline{x}_m}^{(m)} \bigg) \\
& \quad\quad\quad A_{x_1,...,x_M} \bigg( Y_{x_1,...,x_M} \otimes \hat{1} - \sum_{\overline{x}_{m^\prime}^{\prime}} Y_{\overline{x}_{m^\prime}^{\prime}}^{(m^{\prime})} \otimes A_{\overline{x}_{m^\prime}^{\prime}}^{(m^{\prime})} \bigg)^{\dagger} \bigg]  \\
 = &\sum_{x_m,x_{m'}} \Tr\bigg[\bigg( Y^{(m)}_{x_m}\otimes \hat{1} -  \sum_{\overline{x}_m} Y_{\overline{x}_m}^{(m)}\otimes
A_{\overline{x}_m}^{(m)}   \bigg)\\
&  \quad\quad\quad
   A^{(m)}_{x_m}\otimes A^{(m')}_{x_{m'}}  \bigg( Y_{x_{m'}}^{(m')}\otimes \hat{1} -   \sum_{\overline{x}'_{m'}}  Y_{\overline{x}'_{m'}}^{(m')}
\otimes A_{\overline{x}'_{m'}}^{(m')}   \bigg)^{\dagger}\rho \bigg].
\end{split}
\end{equation}
Expanding the last two parentheses  in (\ref{Eq:PmRPmprime}), and using the fact that  $\{A^{(m)}_{x_m}\}_{x_m}$ is a POVM and thus  $\sum_{x_m}  A^{(m)}_{x_m} = \hat{1}$, yields $P_mRP_{m^\prime}= 0$.
Hence, $R$ is block-diagonal with respect to the subspaces $\mathcal{V}_m$, i.e., $R = \sum_{m}P_mRP_m$.
\end{proof}

\subsection{Semidefinite test for quantum networks}

From Supplementary Proposition \ref{Prop:DecomposingQuantumNoInputsAgain} we can conclude that whenever we find a covariance matrix that cannot be decomposed in the manner of (\ref{vdfbvsdfagain}) and (\ref{Eq.ThmProjectorsCAndRagain}), then we can exclude the quantum network $G$ as an explanation of the observed covariance. This observation can thus be turned into a test that allows us to falsify hypothetical quantum networks, and we colloquially refer to this as the `semidefinite test'. In Section \ref{SecWitness} we discuss how this semidefinite test can be cast as a semidefinite  program.

 The semidefinite test provides an outer relaxation to the set of distributions $P$ that are compatible with $G$. In other words, the set of distributions (or covariance matrices) that are compatible with $G$ forms a subset of the distributions (covariance matrices) that are accepted by the semidefinite test. Hence, if a distribution fails the test, we can safely reject $G$ as an explanation, while if the distribution passes the test, it may still not be compatible with $G$.

On the level of the covariance matrix $\mathrm{Cov}(Y)$, the conditions (\ref{vdfbvsdfagain}) and (\ref{Eq.ThmProjectorsCAndRagain}) in the quantum case are identical to the conditions (\ref{DecCovClassical}) and (\ref{DecCovClassical}) in the classical  case. Hence, the semidefinite test in the  quantum case is identical to the test in the classical case, which means that it tests the topology of the network, irrespective of whether this network is classical or quantum.

%%%%%%%%%%%%%%%%%%%%%%%%%%%%%%%%%%%%%%%%%%%%%%%%%%%%%%%%%

\section{\label{SecWitness}Dual formulation and a general witness for quantum networks with bipartite sources}

In this Section we prove some of the claims made in the main text, regarding the results shown in figures 1 and 2. In the following section, we establish an equivalent dual formulation for the test introduced in Proposition 1 of the main text, which has the advantage of providing a witness for the particular graph topology imposed by the causal model under consideration---i.e. it provides a linear functional over the covariance matrix $\mathrm{Cov}(Y)$ that determines incompatibility with the proposed causal model for values above a certain threshold.

\subsection{Dual formulations}

As established in Proposition 1, the decomposition imposed by a quantum network on the covariance matrix $\mathrm{Cov}(Y)$ (which we denote here as $\mathcal{C}$ for convenience) can be immediately stated in the form of a semidefinite program (SDP) as a feasibility problem \cite{cavalcanti2016quantum}, namely
\begin{subequations}
\begin{align}
\text{Given} &\quad \mathcal{C}, \label{eq:app_primal_given}\\
\text{find} &\quad R,\,C_n, \label{eq:app_primal_feasib_obj}\\
\text{subject to} &\quad R\geq 0,\label{eq:app_primal_const_RPsd}\\
&\quad R = \sum_m P_m R P_m, \label{eq:app_primal_const_RStruct}\\
&\quad C_n \geq 0, \label{eq:app_primal_const_CnPsd}\\
&\quad C_n = P^{(n)} C_n P^{(n)}, \label{eq:app_primal_const_CnStruct}\\
&\quad \mathcal{C} = R + \sum_n C_n. \label{eq:app_primal_const_CovDecomp}
\end{align}
\label{eq:app_primal_sdp}
\end{subequations}
For practical purposes, this test can be slightly simplified into 
\begin{subequations}
\begin{align}
\text{Given} &\quad \mathcal{C}, \label{eq:app_primal_simp_given}\\
\text{find} &\quad R,\,C_n, \label{eq:app_primal_simp_feasib_obj}\\
\text{subject to} &\quad R\geq 0, \label{eq:app_primal_simp_const_RPsd}\\
&\quad C_n \geq 0, \label{eq:app_primal_simp_const_CnPsd}\\
&\quad \mathcal{C} = \sum_m P_m\,R\,P_m + \sum_n P^{(n)}\,C_{n}\,P^{(n)},  \label{eq:app_primal_simp_const_CovDecomp}
\end{align}
\label{eq:app_primal_simplified}
\end{subequations}
which subsumes the expected structures for $R$ and $C_n$ [Eqs.\ (\ref{eq:app_primal_const_RStruct}, \ref{eq:app_primal_const_CnStruct})] within the decomposition for $\mathcal{C}$ [Eq.\ \eqref{eq:app_primal_simp_const_CovDecomp}]. There is no loss in this simplification since we may always get a solution for problem \eqref{eq:app_primal_sdp} from the solution for \eqref{eq:app_primal_simplified} and vice-versa: Any solution for the original problem automatically satisfies the constraints of the simplified problem, while, on the other way, if the $R$ and $C_n$ obtained from \eqref{eq:app_primal_simplified} are not already a solution for the original problem, we may define
\begin{align}
R^\prime & \coloneqq \sum_m P_m\, R\, P_m, \\
C^\prime_n & \coloneqq P^{(n)}\, C_n\, P^{(n)},
\end{align}
and the new variables make a viable solution for both \eqref{eq:app_primal_sdp} and \eqref{eq:app_primal_simplified}. In fact, we have $R^\prime \geq 0$ (since $R \geq 0$ from Eq.\ \eqref{eq:app_primal_simp_const_RPsd}), and $C^\prime_n \geq 0$; $\mathcal{C} = R^\prime + \sum_n C^\prime_n$, from Eq.\ \eqref{eq:app_primal_simp_const_CovDecomp}; $\sum_k P_k\, R^\prime\, P_k = R^\prime$, since
\begin{align}
\sum_k P_k\,R^\prime\,P_k &= \sum_{k,m} P_k\, P_m\, R\, P_m\, P_k \nonumber \\
&= \sum_m P_m\, R\, P_m = R^\prime,
\end{align}
given that $P_m\, P_k = \delta_{m,k}\, P_m$. Similarly,
\begin{align}
P^{(n)}\, C^\prime_n\, P^{(n)} &= \sum_{\substack{j,j^\prime \\ k,k^\prime}\, \in\, \mathcal{C}_n} P_j\, P_k\, C_n\, P_{k^\prime}\, P_{j^\prime} \nonumber \\
&= \sum_{k,k^\prime\, \in\, \mathcal{C}_n} P_k\, C_n\, P_{k^\prime} \nonumber \\
&= P^{(n)}\, C_n\, P^{(n)} = C^\prime_n.
\end{align}

From the SDP \eqref{eq:app_primal_simplified}, we may obtain an associated dual optimization problem \cite{boyd2004convex,cavalcanti2016quantum} that returns a numerical value for the optimal separation between the given covariance matrix $\mathcal{C}$ and the set of matrices that satisfy the constraints (\ref{eq:app_primal_simp_const_RPsd}-\ref{eq:app_primal_simp_const_CovDecomp}). More specifically, the test described by Eqs.\ \eqref{eq:app_primal_simplified} admits the dual formulation
\begin{subequations}
\begin{align}
\substack{\textstyle\text{Maximize}\\W} &\quad \Tr[W\,\mathcal{C}], 
\label{eq:app_sdp_dual_objective}\\
\text{subject to} &\quad \sum_m P_m W P_m \leq 0, 
\label{eq:app_sdp_dual_RConst}\\
&\quad P^{(n)} W P^{(n)} \leq 0,
\label{eq:app_sdp_dual_CnConst}
\end{align}
\label{eq:app_sdp_dual}
\end{subequations}
which should detect feasibility or infeasibility in the same way that the other (primal) formulation does. As a byproduct, we obtain the hermitian matrix $W$, which works as a witness for incompatibility with the the causal model under test: the dual constraints (\ref{eq:app_sdp_dual_RConst}-\ref{eq:app_sdp_dual_CnConst}) ensure that $\Tr[W\,\mathcal{C}]\leq 0$ whenever $\mathcal{C}$ is decomposable as $R~+~\sum_n C_n$, since 
\begin{align}
\Tr[W\,R] &= \sum_m \Tr[W\,P_m R P_m] \nonumber \\
&= \sum_m\Tr[P_m W P_m\,R] \leq 0,
\end{align}
and similarly $\Tr[W\,C_n] \leq 0$. As a consequence, if $W$ is applied over another given covariance matrix $\mathcal{C}^\prime$ and $\Tr[W\,\mathcal{C}^\prime] > 0$, we can immediately conclude that $\mathcal{C}^\prime$ is not explained by the proposed causal model.

While formulation \eqref{eq:app_sdp_dual} is very simple and already useful for obtaining a valid witness for the given causal model, its objective function may be unbounded in some cases, where the primal problem is infeasible. It is interesting then for practical purposes to have the witness $W$ bounded in some manner. To this end, we add a scaling constraint on $\Tr[W]$, rendering the program as
\begin{subequations}
\begin{align}
\substack{\textstyle\text{Maximize}\\W} &\quad \Tr[W\,\mathcal{C}], \label{eq:app_dual_sc_obj}\\
\text{subject to} &\quad \sum_m P_m W P_m \leq 0, \label{eq:app_dual_sc_const1}\\
&\quad P^{(n)} W P^{(n)} \leq 0, \label{eq:app_dual_sc_const2}\\
&\quad \Tr[W] \geq -1, \label{eq:app_dual_sc_scale}
\end{align}
\label{eq:app_sdp_dual_scaled}
\end{subequations}
noting that $\Tr[W]$ is already bounded above due to Eq.\ \eqref{eq:app_dual_sc_const1}, since $\Tr[W] = \Tr[\,\sum_m P_m\, W] = \sum_m \Tr[P_m\, W\, P_m] \leq 0$, using that $\sum_m P_m = \mathbbm{1}$ and $P^2_m = P_m$.
 
\subsection{Applications} 

As an application, we consider the family of distributions
\begin{equation}
P^{N}_{pq}(x_1,..,x_N) = p\,\delta^{(N)}_{0} + q\,\delta^{(N)}_{1} + (1-p-q)\,\frac{1 - \delta^{(N)}_{0} - \delta^{(N)}_{1}}{2^N -2}
\label{eq:app_pq_dist_general}
\end{equation}
for $N$ parties with binary outcomes, where $\delta^{(N)}_{x}=1$ if $x_1=x_2=...=x_N=x$, and $0$ otherwise, and $p+q\leq 1$. These distributions include, in particular, the ones used for the results in Figs. 1-c and 2-c, which correspond, respectively, to the cases $N=3$ and $N=4$. Figures 1-c and 2-c also include the results based on the Finner inequality, that in triangle case can be simply written as $p(x_1,x_2,x_3) \leq \sqrt{p(x_1)p(x_2)p(x_3)}$ (see \cite{Renou2019} for more details).

To build the covariance matrix, we use the feature maps $Y^{(m)}_{x_m} = |x_m\rangle$, where $\{\,|x_m\rangle\,\vert\, 1 \leq x_m \leq X_m\,\}$ forms the canonical basis in the $X_m$-dimensional Hilbert space $\mathcal{V}_m$. Let now $a_d$ be a ladder operator in $d$ dimensions defined as $a_d \coloneqq \sum_{j=1}^{d-1} | j+1 \rangle\langle j |$, $\mathbbm{1}_{d}$ the identity also in $d$ dimensions, and $\sigma_x,\sigma_y,\sigma_z$ the usual Pauli matrices. Numerical evidence suggests that the witness
\begin{equation}
W_{2N} \coloneqq -\mathbbm{1}_{2N} + 2\,\left(\mathbbm{1}_N \otimes \sigma_x\right) + \sum_{j=1}^{2N-1} (-1)^j\left[ a^{\,j}_{2N} + \left(a^{\dagger}_{2N}\right)^j\right]
\label{eq:general_witness}
\end{equation}
is optimal for the family of distributions $P^{N}_{pq}$ when the corresponding covariance matrix is incompatible with a network where all possible bipartite sources (i.e., sources connecting 2 children) are present (See, for instance, Figures 1-a and 2-a in the main text for the cases $N=3$ and $N=4$). This has been verified numerically for $N=3,...,7$ parties and we conjecture that this witness is always optimal for this family of distributions. Nonetheless, regardless of optimality, it is possible to prove that $W_{2N}$ is always a valid witness and, in particular, that for any value of $p$, there is always a value of $q$ above which the witness detects incompatibility with the causal model. 

To prove that $W_{2N}$ is a valid witness, first we use that, for the scenario that we consider here, where all parties have binary outcomes, the projector onto $\mathcal{V}_m$ can be written as $P_m = |m\rangle\langle m|\otimes\mathbbm{1}_2$. We note also that $a_{2N}~=~\mathbbm{1}_N \otimes a_2 + a_N \otimes a^\dagger_2$.
Together these imply that
\begin{align}
P_m\, a_{2N}^{k}\, P_m   &= \delta_{k,1}\,|m\rangle\langle m| \otimes a_2,\\
P_m\, a_{2N}^{2k}\, P_n   &= \delta_{m,n+k}\,|m\rangle\langle n| \otimes \mathbbm{1}_2,\\
P_m\, a_{2N}^{2k+1}\, P_n &= \delta_{m,n+k}\,|m\rangle\langle n| \otimes a_2 + \delta_{m,n+k+1}\,|m\rangle\langle n| \otimes a_2^{\dagger},
\end{align}
for $m, n \in \{1,...,N\}$ and $m \neq n$. Applying these results on Eq.\ \eqref{eq:general_witness}, we obtain 
\begin{equation}
P_m\, W_{2N}\,P_n = (-1)^{\delta_{m,n}}\,|m\rangle\langle n|\otimes (\mathbbm{1}_2 - a_2  - a_2^\dagger) = (-1)^{\delta_{m,n}}\,|m\rangle\langle n|\otimes|-\rangle\langle -|,
\end{equation}
where $|-\rangle \coloneqq |0\rangle - |1\rangle$. Consequently, 
\begin{equation}
\sum_m P_m\,W_{2N}\,P_m = -\mathbbm{1}_N \otimes |-\rangle\langle -|\, \leq 0,
\end{equation} 
and, given that each latent vertex connects only two different parties, 
\begin{equation}
P^{(n)}\,W_{2N}\,P^{(n)} = (P_{i_n} + P_{j_n})\,W_{2N}\,(P_{i_n} + P_{j_n}) = -\left(|-\rangle\langle -|_{i_n,j_n} \otimes |-\rangle\langle -|\right) \leq 0,
\end{equation}
where $|-\rangle_{ij} \coloneqq |i\rangle - |j\rangle$ for $i\neq j$, and $P_{i_n},\,P_{j_n}$ are the projectors that compose $P^{(n)}$, $i_n \neq j_n$. Consequently, $W_{2N}$ satisfy the dual constraints \eqref{eq:app_dual_sc_const1} and \eqref{eq:app_dual_sc_const2} and, therefore, returns $\Tr[W_{2N}\,\mathcal{C}] \leq 0$ for any covariance matrix that is compatible with the proposed causal model.

\begin{figure}[htb]
    \centering
    \includegraphics[width=0.45\textwidth]{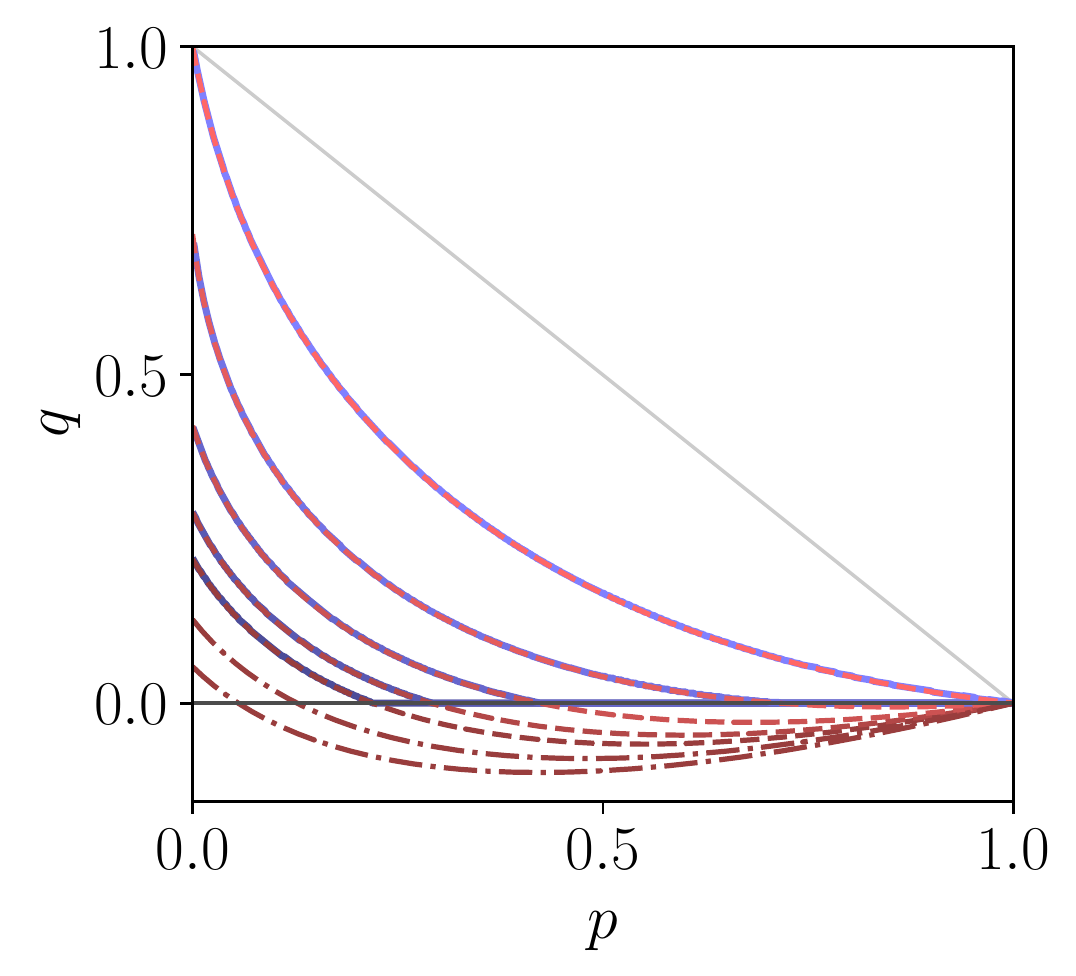}
    \caption{Comparison between numerical results (solid blue lines) obtained as solutions of the SDP \eqref{eq:app_sdp_dual_scaled} and the analytical curves (dashed red lines) that bound the region defined in \eqref{eq:app_analytic_form}. Corresponding numerical and analytic solutions are presented for $N=3,...,7$, where it can be observed that both types of solution coincide, implying that the witness $W_{2N}$ [Eq.\ \eqref{eq:general_witness}] is optimal for the entire family of distributions $P^N_{pq}$ in these cases. The dot-dashed curves correspond to analytical curves for $N=10$ and $N=20$.}
    \label{fig:comparison_analytic}
\end{figure}

Now, let $C^{N}_{pq}$ be the covariance matrix for $P^N_{pq}$. It can be written as
\begin{equation}
C^N_{pq} = \left(\Delta^N_{pq}\,\mathbbm{1}_N + \chi^N_{pq}\,|\mathbf{1}\rangle\langle \mathbf{1}|_N\right)\otimes (\mathbbm{1}_2 + \sigma_x),
\label{eq:app_cov_pqdist}
\end{equation}
where $\Delta^N_{pq} \coloneqq 2^{N-2}(1-p-q)/(2^N-2)$, $\chi^N_{pq} \coloneqq \frac{1}{4}[1 - (p-q)^2] - \Delta^N_{pq}$, and $|\mathbf{1}\rangle_N \coloneqq \sum_{m=1}^N |m\rangle$. Evaluating $\Tr[W_{2N}\,C^N_{pq}]$, we obtain the expression $\Tr[W_{2N}\,C^N_{pq}] = 4N\,[\chi^N_{pq}(N-2) - \Delta^N_{pq}]$. This allows us to solve for $q$ and establish the analytic formula 
\begin{equation}
q > p + \kappa_N - \sqrt{4\,\kappa_N\, p + (\kappa_N -1)^2},
\label{eq:app_analytic_form}
\end{equation}
for the region of incompatibility with the network witnessed by $W_{2N}$ (i.e., the region where $\Tr[W_{2N}\,C^N_{pq}] > 0$ ), where $\kappa_N \coloneqq \frac{(N-1)\,2^{N-2}}{(N-2)\,(2^{N-1}-1)}$. 

It should be remarked that, since $\chi^N_{pq}$ and $\Delta^N_{pq}$ are both symmetric under permutation of $p$ and $q$, the same is true for the region described by \eqref{eq:app_analytic_form}, which is expected since $C^N_{pq}$ has this property. Also, if equality is obtained in Eq.\ \eqref{eq:app_analytic_form} for $q = q_0 < 1$ at $p=0$, then $C^N_{pq}$ is always incompatible with the network independently of $p$ for $q>q_0$, with the same behavior occurring for $p > q_0$, due to the symmetry. In fact, since $q_0 = \kappa_N - |\kappa_N - 1|$, this always true for $N \geq 4$, where $\kappa_N < 1$. As $N \rightarrow +\infty$, $\kappa_N \rightarrow 1/2$, implying then that $q_0 \rightarrow 0^+$. In fig. \ref{fig:comparison_analytic} we show the analytical curves for different values of $N$ and a comparison with the numerical results.

%%%%%%%%%%%%%%%%%%%%%%%%%%%%%%%%%%%%%%%%%%%%%%%%%%%%%%%%%%

\section{\label{SecClassicalWithInputs}  Semidefinite decompositions for classical networks with inputs}

Here we combine the classical test developed in \cite{KelaEtAl20} (see Section \ref{SecClassicalWithoutInputs}) with the notion of `inputs' or `measurement settings'.  We do this for the sake of comparison with the quantum test with inputs in Proposition 2, which we prove in Section \ref{SecProofOfBipartWithChoice}.
As we shall see here, the notion of inputs does in the classical case not introduce anything essentially new compared to \cite{KelaEtAl20}, in the sense that the classical setting with inputs can be simulated by an `extended' classical setting without inputs.

For a DAG $G$ with parents $p_1,\ldots,p_N$ and children $c_1,\ldots,c_M$ with inputs or (measurement) settings $\overline{s} = (s_1,\ldots,s_M) \in \{1,\ldots,S_1\}\times\cdots\times\{1,\ldots S_M\}$, the joint distribution of the parents and children, conditioned on the inputs, can be written
\begin{equation}
\label{ClassicalWithInputs}
\begin{split}
P(c_1,\ldots,c_M,p_1,\ldots,p_N|s_1,\ldots,s_M) = & \Pi_{m=1}^{M}P(c_m|s_m,\mathcal{P}_m)\Pi_{n=1}^{N}P(p_n).
\end{split}
\end{equation}
One may note that (\ref{ClassicalWithInputs}) is nothing but the ordinary causal Markov condition, applied to the scenario with inputs.
 It expresses that each child $c_m$ may potentially be influenced, not only by its parents $\mathcal{P}_m$, but also by the choice of input $s_m$. One should in particular note that each input $s_m$ only affects the output of child $c_m$. We should also remark that we here  overload the notion and let $p_1,\ldots,p_N$ and $c_1,\ldots,c_M$ not only denote vertices in the DAG $G$, but also the random variables associated with those vertices, as well as the values that these random variables may attain.
 
We say that an observed conditional  distribution $P(c_1,\ldots,c_M|s_1,\ldots,s_M)$ is compatible with a given classical causal structure $G$ with parents $p_1,\ldots,p_N$ and children $c_1,\ldots,c_M$, with associated inputs $s_1,\ldots,s_M$, if it is the marginal of some distribution of the form (\ref{ClassicalWithInputs}), for all inputs $(s_1,\ldots,s_M) \in \{1,\ldots,S_1\}\times\cdots\times\{1,\ldots S_M\}$.

\subsection{\label{SecProofIdea} Proof-idea and conceptual remarks}

Before turning to the statement and proof of Supplementary Proposition \ref{PropClassicalWithInputs} below, we will in the following give an outline of the main conceptual ideas of our argument, which in essence reduces the proof to a special case of Supplementary  Proposition \ref{PropClassicalWithoutInputs}. 
Suppose that we have a (classical) device that randomly produces an output $X$, but in such a way that this distribution is determined by some input $s = 1,\ldots,S$. In other words,  the output distribution $P(X = x|s)$ is conditioned on $s$. Now imagine a second device that  simultaneously produces a collection of random variables  $X^{(1)},\ldots,X^{(S)}$, with a joint distribution $P(X^{(1)},\ldots X^{(S)})$, for which the marginal distribution of $X^{(s)}$ is $P(X^{(s)} = x)  = P(X = x|s)$. There always exists at least one such distribution, namely the product distribution $P(X^{(1)} = x^{(1)},\ldots, X^{(S)} = x^{(S)}) = \Pi_{s=1}^{S}P(X^{(s)} = x^{(s)}|s)$.  Marginalizing over all indices but $s$, the second device can thus simulate the first by presenting only random variable $X^{(s)}$ when the input is $s$.  In the proof of Supplementary Proposition \ref{PropClassicalWithInputs}, we will apply this `extension' to all the children (that has more that one measurement setting $s_m$). 

As a side-remark, one may note that for the corresponding quantum case,  in Proposition 2, the parents potentially distribute entangled states to their children. Due to such entangled states, it is difficult to see that one could employ a similar proof strategy of `extensions' in the quantum case. Hence, although the quantum version in Proposition 2 shares an essential structural similarity with Supplementary Proposition \ref{PropClassicalWithInputs}, the proof strategy of the former is very different from the latter, and focuses directly on the covariance matrix.

We next make a few observations and definitions concerning the framework of Supplementary Proposition \ref{PropClassicalWithInputs}.
We consider a scenario where each child $c_m$ at each instance only can choose one single  input $s_m$. For each child $c_m$ and for each choice of local input $s_m$ we assign a feature map $Y^{(m,s_m)} := Y^{(m,s_m)}(c_m)$ into a subspace $\mathcal{V}_{m,s_m}$.  For each single child, we can observe $\mathrm{Cov}(Y^{(m,s_m)}(c_m)|s_m)$. Moreover, we can observe all kinds of cross covariances  $\mathrm{Cov}(Y^{(m,s_m)}(c_m),Y^{(m',s'_{m'})}(c_{m'})|s_m,s'_{m'})$ for each possible pair of inputs $s_m, s'_{m'}$ of child $m,m'$ with $m\neq m'$. Additionally, one may note that $\mathrm{Cov}(Y^{(m,s_m)}(c_m)|\overline{s})= \mathrm{Cov}(Y^{(m,s_m)}(c_m)|s_m)$  and $\mathrm{Cov}\big(Y^{(m,s_m)}(c_m),Y^{(m',s'_{m'})}(c_{m'})\big|\overline{s}\big) =  \mathrm{Cov}\big(Y^{(m,s_m)}(c_m),Y^{(m',s'_{m'})}(c_{m'})\big|s_m,s'_{m'}\big)$,  due to the structure of (\ref{ClassicalWithInputs}).
 However, in this setting, there exist no cross covariances of the type  `$\mathrm{Cov}\big(Y^{(m,s_m)}(c_m),Y^{(m,s'_{m})}(c_m)\big)$' for $s'_m\neq s_m$, since at each instance, $c_m$ can only choose one input $s_m$.  We collect all the observable covariances into  the \emph{observable covariance matrix}   
\begin{equation}
\label{ClassicalObservableCovariance}
\begin{split}
C_{\mathrm{observable}} := &   \sum_{m}\sum_{s_m}\mathrm{Cov}(Y^{(m,s_m)}(c_m)|s_m) \\
&  +\sum_{\substack{m,m' \\ m\neq m'}}\sum_{s_m,s'_{m'}} \mathrm{Cov}\big(Y^{(m,s_m)}(c_m),Y^{(m',s'_{m'})}(c_{m'})\big|s_m,s'_{m'}\big).
\end{split}
\end{equation}
We wish to find a `completion' $C_{\mathrm{completion}}$  of the observable  covriances $C_{\mathrm{observable}}$, such that their sum, $C := C_{\mathrm{observable}} + C_{\mathrm{completion}}$, is positive semidefinite, and where $C_{\mathrm{completion}}$ satisfies.
\begin{equation}
\label{CompletionRequirements}
\begin{split}
& P_{m,s_m}C_{\mathrm{completion}}P_{m,s_m} = 0,\quad  m = 1,\ldots,M,\quad  s_m = 1,\ldots,S_m\\
& P_{m,s_m}C_{\mathrm{completion}}P_{m',s'_{m'}} = 0,\quad    m ,m' = 1,\ldots,M: m\neq m',\quad  s_m =1,\ldots, S_m,\quad s'_{m'}= 1,\ldots, S_{m'}.
\end{split}
\end{equation}
These conditions in essence says that $C_{\mathrm{completion}}$, regarded as a block matrix with respect to the sub spaces $\mathcal{V}_{m,s_m}$, must be zero on the blocks that constitutes $C_{\mathrm{observable}}$.
 In the proof of Supplementary Proposition \ref{PropClassicalWithInputs} this completion is constructed via the above described `extended' random variables. The point is that the cross covariances $\mathrm{Cov}\big(Y^{(m,s_m)}(c_m),Y^{(m,s'_{m})}(c_m)\big)$ that lack meaning in the conditional setting, become meaningful in the extended setting. Moreover, via the extended collection of random variables, it turns out that Supplementary Proposition \ref{PropClassicalWithoutInputs} can be applied to to the `completed' matrix $C$, which  yields Supplementary Proposition \ref{PropClassicalWithInputs}.

\subsection{Proof of Supplementary Proposition \ref{PropClassicalWithInputs}}

\begin{Proposition}
\label{PropClassicalWithInputs}
Let the conditional distribution $P(c_1,\ldots,c_M|s_1,\ldots,s_M)$ be compatible with the classical causal structure $G$ with parents $p_1,\ldots,p_N$ and children $c_1,\ldots,c_M$, with associated inputs $s_1,\ldots,s_M$.  Assume that each child  $c_m$, and each input $s_m$, is assigned a feature map $Y^{(m,s_m)}$ into a vector space $\mathcal{V}_{m,s_m}$. 
Let the operator $C_{\mathrm{observable}}$ on $\mathcal{V} := \bigoplus_{m=1}^{M}\bigoplus_{s_m=1}^{S_m}\mathcal{V}_{m,s_m}$  be as  defined in (\ref{ClassicalObservableCovariance}).
 Then there exist operators $C_{\mathrm{completion}}$, $\tilde{R}$ and $(\tilde{C}_n)_{n=1}^{N}$ on $\mathcal{V}$, such that $C_{\mathrm{completion}}$  satisfies (\ref{CompletionRequirements}), and 
\begin{equation}
\label{ConditionCompletionClassical}
C:= C_{\mathrm{observable}} + C_{\mathrm{completion}}\geq 0,
\end{equation}
and
\begin{eqnarray}
\label{sfgbsfgbClassical}
C & = & \tilde{R} + \sum_{n=1}^{N}\tilde{C}_n,\quad \tilde{R}\geq 0,\quad \tilde{C}_n\geq 0,\\
\label{fgndnfhd}
& & \tilde{P}^{(n)}\tilde{C}_n \tilde{P}^{(n)} = \tilde{C}_n,\quad \tilde{R} = \sum_{m,s_m}P_{m,s_m}\tilde{R}P_{m,s_m},\\
\label{mhdghmggdClassical}
& & \tilde{P}^{(n)}:= \sum_{m\in\mathcal{C}_n}\sum_{s_m=1}^{S_m}P_{m,s_m},
\end{eqnarray}
 where $P_{m,s_m}$ is the projector onto $\mathcal{V}_{m,s_m}$, and where $\mathcal{C}_n$ denotes the children having parent $p_n$ in the given DAG $G$.
\end{Proposition}
\emph{Remark:} One may note that from the requirement that $C$ is positive semidefinite, it follows that $C$ also is Hermitian, $C^{\dagger} = C$. Since $C_{\mathrm{observable}}$ necessarily is Hermitian, it follows that $C_{\mathrm{completion}} = C -C_{\mathrm{observable}}$ also is Hermitian. Hence, in any  search for a completion  $C_{\mathrm{completion}}$, one can, without loss of generality, assume that $C_{\mathrm{completion}}^{\dagger} = C_{\mathrm{completion}}$.

\begin{proof}
By the asumption that $P(c_1,\ldots,c_M|s_1,\ldots,s_M)$ it compatible with the classical causal structure $G$ with parents $p_1,\ldots,p_N$ and children $c_1,\ldots,c_M$, with associated inputs $s_1,\ldots,s_M$, it follows that $P(c_1,\ldots,c_M|s_1,\ldots,s_M)$ is the margin of a conditional joint distribution $P(c_1,\ldots,c_M,p_1,\ldots,p_N|s_1,\ldots,s_M)$, as in (\ref{ClassicalWithInputs}), with respect to $G$.

 The first step is to construct a new graph $\tilde{G}$ that contains the same set of parents as $G$, but where each child $c_m$ in $G$ is replaced with a collection of  children $c_m^{(1)},\ldots, c_m^{(S_m)}$, i.e., one new child for each input $s_m$. 
 We moreover assume that the set of parents of child $c_m^{(s_m)}$ is $\mathcal{P}_m$. In other words, child $c_m^{(s_m)}$  in $\tilde{G}$ has the same set of parents as $c_m$ has in $G$. 
 
 In the following we  construct an extended probability distribution $\tilde{P}$ of  $\{c_{m}^{(s_m)}\}_{m,s_m}$, $p_1,\ldots, p_N$ regarded as  random variables.
The distribution $P(c_1,\ldots,c_M,p_1,\ldots,p_N|s_1,\ldots,s_M)$ is by assumption  of the form (\ref{ClassicalWithInputs}), which thus yields a collection of conditional distributions $P(c_m|s_m,\mathcal{P}_m)$.  
Define $P(c^{(s_m)}_m |\mathcal{P}_m)$ by
\begin{equation}
P(c^{(s_m)}_m = x|\mathcal{P}_m) := P(c_m = x|s_m,\mathcal{P}_m),\quad m = 1,\ldots,M,\quad s_m = 1,\ldots, S_m,
\end{equation}
and construct the distribution of $p_1,\ldots,p_N, \{c_{m}^{s_m}\}_{m,s_m}$ by
\begin{equation}
\label{gbfdgn}
\begin{split}
\tilde{P}(c_1^{(1)},\ldots,c_{1}^{(S_1)},\ldots,c_M^{(1)},\ldots,c_M^{(S_M)},p_1,\ldots,p_N) = & 
\Pi_{m=1}^{M}\Pi_{s_m=1}^{S_m}P(c_m^{(s_m)}|\mathcal{P}_m)\Pi_{n=1}^{N}P(p_n).
\end{split}
\end{equation}
If we take the marginal distribution of $c_1^{(1)},\ldots,c_{1}^{(S_1)},\ldots,c_M^{(1)},\ldots,c_M^{(S_M)}$ in (\ref{gbfdgn}), it follows by comparison of  (\ref{gbfdgn}) with (\ref{ClassicalWithoutInputs}) that the resulting  distribution $\tilde{P}(c_1^{(1)},\ldots,c_{1}^{(S_1)},\ldots,c_M^{(1)},\ldots,c_M^{(S_M)})$ is compatible with the new DAG $\tilde{G}$. Thus Supplementary Proposition \ref{PropClassicalWithoutInputs} is applicable to $\tilde{P}$ with respect to $\tilde{G}$. For child $c_m^{(s_m)}$ we choose the feature map $Y_{m,s_m}$. 
By  Supplementary Proposition \ref{PropClassicalWithoutInputs}, there exist operators $\tilde{R}$ and $(\tilde{C}_n)_{n=1}^{N}$, such that  the covariance matrix $\mathrm{Cov}(\tilde{Y})$ of
\begin{equation}
\tilde{Y}:=\sum_{m=1}^{M}\sum_{s_m=1}^{S_m}Y^{(m,s_m)}(c^{(s_m)}_m),
\end{equation}
 satisfies
\begin{equation}
\label{CovDec}
\mathrm{Cov}(\tilde{Y}) = \tilde{R} + \sum_{n=1}^{N}\tilde{C}_{n},\quad \tilde{R}\geq 0,\quad \tilde{C}_n \geq 0,
\end{equation}
where
\begin{equation}
\label{CondtildeCtildeR}
\begin{split}
&\tilde{P}^{(n)}\tilde{C}_n\tilde{P}^{(n)} = \tilde{C}_n,\quad \tilde{R} = \sum_{m=1}^{M}\sum_{s_m=1}^{S_m}P_{m,s_m} R P_{m,s_m},
\end{split}
\end{equation}
and
\begin{equation}
\label{projdec}
\begin{split}
& \tilde{P}^{(n)}:= \sum_{(m,s_m)\in\mathcal{C}_ n}P_{m,s_m} = \sum_{m\in\mathcal{C}_ n}\sum_{s_m=1}^{S_m}P_{m,s_m},
\end{split}
\end{equation}
where in the second equality in (\ref{projdec}) we have used the observation that since $c_m^{(s_m)}$ in $\tilde{G}$ has the same parents as $c_m$ in $G$, it follows that if $c_m$ is a child with parent $p_n$ in $G$, then all of $c_m^{(1)},\ldots,c_m^{(S_m)}$ are children with parent $p_n$ in $\tilde{G}$.

In the following, we  compare  marginal distributions of (\ref{gbfdgn}) with those of (\ref{ClassicalWithInputs}). The marginal distribution of  $c_m^{(s_m)},p_1,\ldots,p_N$, with respect to  (\ref{gbfdgn}), can be written
\begin{equation}
\begin{split}
\tilde{P}(c_m^{(s_m)},p_1,\ldots,p_N) = & P(c_m^{(s_m)}|\mathcal{P}_m)\Pi_{n=1}^{N}P(p_n).
\end{split}
\end{equation}
Similarly, if we marginalize the conditional distribution (\ref{ClassicalWithInputs}) to $c_m,p_1,\ldots,p_N$, we obtain
\begin{equation}
\begin{split}
P(c_m,p_1,\ldots,p_N|s_1,\ldots,s_M) = &P(c_m|s_m,\mathcal{P}_m)\Pi_{n=1}^{N}P(p_n),
\end{split}
\end{equation}
where we note that only the conditioning on $s_m$ remains on the right hand side.
By recalling that we by construction have $P(c_m^{(s_m)} = x |\mathcal{P}_m) = P(c_m = x|s_m,\mathcal{P}_m)$, we can conclude that
\begin{equation}
\label{adfbdfb}
\begin{split}
\tilde{P}(c_m^{(s_m)}=x,p_1,\ldots,p_N)  = P(c_m = x,p_1,\ldots,p_N|s_m),
\end{split}
\end{equation}
where we on the right hand side have dropped all superfluous conditioning. 

Analogously, for $m\neq m'$ we find that the marginalization   of (\ref{gbfdgn}) to $c_m^{(s_m)},c_{m'}^{(s'_{m'})}, p_1,\ldots,p_N$ yields
\begin{equation}
\begin{split}
\tilde{P}(c_m^{(s_m)},c_{m'}^{(s'_{m'})},p_1,\ldots,p_N) = & 
P(c_m^{(s_m)}|\mathcal{P}_m)P(c_{m'}^{(s'_{m'})}|\mathcal{P}_{m'})\Pi_{n=1}^{N}P(p_n).
\end{split}
\end{equation}
 Similarly, the marginalization of the conditional distribution (\ref{ClassicalWithInputs}) to $c_m,c_{m'},p_1,\ldots,p_N$ yields
\begin{equation}
\begin{split}
P(c_m,c_{m'},p_1,\ldots,p_N|s_1,\ldots,s_M) = &P(c_m|s_m,\mathcal{P}_m)P(c_{m'}|s'_{m'},\mathcal{P}_m)\Pi_{n=1}^{N}P(p_n),
\end{split}
\end{equation}
and  since we again have $P(c_m^{(s_m)} = x |\mathcal{P}_m) = P(c_m = x|s_m,\mathcal{P}_m)$, it follows that 
\begin{equation}
\label{afdbqeb}
\begin{split}
\tilde{P}(c_m^{(s_m)} = x,c_{m'}^{(s'_{m'})} =x',p_1,\ldots,p_N) = P(c_m= x,c_{m'} =x',p_1,\ldots,p_N|s_m,s'_{m'}).
\end{split}
\end{equation}
By (\ref{adfbdfb}) we can conclude that 
\begin{equation}
\label{rznztjmtzj}
\begin{split}
P_{m,s_m}\mathrm{Cov}(\tilde{Y})P_{m,s_m} = & \mathrm{Cov}\big(Y^{(m,s_m)}(c^{(s_m)}_m)\big)\\
= & \mathrm{Cov}\big(Y^{(m,s_m)}(c_m)\big|s_m\big).
\end{split}
\end{equation}
By (\ref{afdbqeb}) it similarly follows that 
\begin{equation}
\label{mizsfgbsfgb}
\begin{split}
P_{m,s_m}\mathrm{Cov}(\tilde{Y})P_{m',s'_{m'}}  =  & \mathrm{Cov}\Big(Y^{(m,s_m)}(c^{(s_m)}_m), Y^{(m',s'_{m'})}(c^{(s'_{m'})}_{m'})  \Big)\\
  =  & \mathrm{Cov}\Big(Y^{(m,s_m)}(c_m), Y^{(m',s'_{m'})}(c_{m'}) \Big|s_m,s'_{m'} \Big).
\end{split}
\end{equation}
Equations (\ref{rznztjmtzj}) and (\ref{mizsfgbsfgb}) implies  that  $C_{\mathrm{observable}}$ defined in (\ref{ClassicalObservableCovariance}) can be written 
\begin{equation}
C_{\mathrm{observable}} =    \sum_{m}\sum_{s_m}P_{m,s_m}\mathrm{Cov}(\tilde{Y})P_{m,s_{m}}
 +\sum_{\substack{m,m' \\ m\neq m'}}\sum_{s_m,s'_{m'}}  P_{m,s_m}\mathrm{Cov}(\tilde{Y})P_{m',s'_{m'}}.
\end{equation}
Moreover, define 
\begin{equation}
C_{\mathrm{completion}} : = \sum_{m=1}^{M}\sum_{\substack{  s_m,s'_m = 1 \\ s_m\neq s'_m}}^{S_m} P_{m,s_m}\mathrm{Cov}(\tilde{Y})P_{m,s'_{m'}}.
\end{equation}
One can confirm that $C_{\mathrm{completion}}$ so defined satisfies all the conditions in (\ref{CompletionRequirements}). 
Moreover,
\begin{equation}
\label{aefbefb}
C_{\mathrm{observable}}  + C_{\mathrm{completion}}  = \mathrm{Cov}(\tilde{Y}).
\end{equation}
Finally, since $ \mathrm{Cov}(\tilde{Y})$ necessarily is positive semidefinite, we can conclude that  $C_{\mathrm{completion}}$ and  $C:=\mathrm{Cov}(\tilde{Y})$ satisfy (\ref{ConditionCompletionClassical}).
By combining (\ref{aefbefb}) with (\ref{CovDec}), (\ref{CondtildeCtildeR}) and (\ref{projdec}), it follows that 
(\ref{sfgbsfgbClassical}), (\ref{fgndnfhd}) and (\ref{mhdghmggdClassical}) holds, which proves the proposition.
\end{proof}

%%%%%%%%%%%%%%%%%%%%%%%%%%%%%%%%%%%%%%%%%%%%%%%%%%%%%%%%%

\section{\label{SecProofOfBipartWithChoice} Proof of Proposition 2 in the main text}

Here we consider the classes of  quantum networks where we have the freedom to locally choose among  collections of measurement settings. In other words, analogous to how Supplementary Proposition \ref{PropClassicalWithInputs} generalizes Supplementary Proposition \ref{PropClassicalWithoutInputs} to the case of inputs, Proposition 2 generalizes Proposition 1 to the case when we have local measurement settings.

Like for the classical case in Section \ref{SecClassicalWithInputs}, completions of the observable covariance matrix is an important notion also in the quantum setting. In the classical case, we made the completion on the level of probability  distributions, where the completion of the covariance matrix followed as a consequence. As already remarked in Section \ref{SecClassicalWithInputs}, it is difficult to see how one could apply a similar proof-strategy in the quantum case, since the quantum parents may generate entanglement between their children. Locally copying the quantum states, in a manner analogous to how we copy the transition probabilities in the classical case,  is thus excluded due to the no-cloning theorem. We circumvent such issues by instead directly completing the covariance matrix itself, and this  approach is detailed in Section \ref{PosDefComplQuantum}.

Let us recall that we here, like in the rest of this investigation, exclusively consider the class of DAGs where the vertices can be partitioned into a class of parents  $p_1,\ldots,p_N$ and a class of children $c_1,\ldots,c_M$, where there only are arrows from parents to children.
For a graph $G$ with parents $p_1,\ldots,p_N$ and children $c_1,\ldots,c_M$, we do, precisely as in Section \ref{SecProofProp1}, associate a structure of Hilbert spaces as in (\ref{Structure}). As in Section \ref{SecProofProp1} we also assume that the parents are initially uncorrelated, with a global state
\begin{equation}
\label{DAGGstatesAgain}
\rho:= \rho_{p_1}\otimes\cdots\otimes\rho_{p_N}\in\mathcal{S}(\mathcal{H}).
\end{equation} 
However, as a generalization of the setting in Section \ref{SecProofProp1}, we here consider a collection of POVM $\{A^{(m,s_m)}_{x_m}\}_{x_m}$,  for $s_m = 1,\ldots, S_m$, where the latter is the possible `inputs' or `measurement settings', resulting in the joint POVMs
 \begin{equation}
 \label{DAGGPOVMinputs}
A_{x_1,\ldots,x_M}^{s_1,\ldots,s_M} := A^{(1,s_1)}_{x_1}\otimes\cdots\otimes A^{(M,s_M)}_{x_M}.
\end{equation}
 This construction results in a distribution of measurement outcomes $x_1,\ldots,x_M$ conditioned on the choice of settings $s_1,\ldots,s_M$, defined by 
\begin{equation}
\label{QuantumWithInputsAgain}
P(x_1,\ldots,x_M|s_1,\ldots,s_M) = \Tr([A^{(1,s_1)}_{x_1}\otimes\cdots\otimes A^{(M,s_M)}_{x_M}][\rho_{p_1}\otimes\cdots\otimes\rho_{p_N}]).
\end{equation}
We say that an observed conditional distribution $P(x_1,\ldots,x_M|s_1,\ldots,s_M)$ is compatible with the given quantum casual structure $G$ with inputs, if it can be written as in (6), for some state as in  (\ref{DAGGstatesAgain}) and some collections of POVMs.

Analogous to the classical case with inputs in Section \ref{SecClassicalWithInputs}, we assign a feature map $Y^{(m,s_m)}$ to each child $c_m$ and each measurement setting $s_m$, i.e., to each measurement outcome $x_{m,s_m}$, corresponding to POVM element $A^{(m,s_m)}_{x_{m,s_m}}$, we associate an element $Y^{(m,s_m)}_{x_{m,s_m}}$ in a vector space $\mathcal{V}_{m,s_m}$.
 We also define $\mathcal{V} := \bigoplus_{m=1}^{M}\bigoplus_{s_m=1}^{S_m}\mathcal{V}_{m,s_m}$. To each subspace $\mathcal{V}_{m,s_m}$ we associate the projector $P_{m,s_m}$. 
 
As a side remark, we note that the central idea in our construction to deal with networks with inputs (Proposition 2) is indeed the same as in the NPA hierarchy \cite{NPA1}. To illustrate, consider the simple bipartite Bell scenario (with a single source). In a Bell experiment we have access to terms like $\Tr(A^{(1,s_1)}_{x_1}A^{(2,s_2)}_{x_2}\rho)$, but non-observable terms like  $\Tr(A^{(1,s_1)}_{x_1}A^{(1,s^{\prime}_1)}_{x^{\prime}_1}\rho)$ are also mathematically well defined. If the observed correlations have a quantum realization,  then it is possible to complete these values in such a way that a larger matrix collecting these extra terms is positive semi-definite. 

The same idea is behind our proof. In fact, we notice that for a single source, our SDP test is equivalent to the first level in the NPA hierarchy. However, for more two or more sources, there is no direct comparison between our method and the NPA hierarchy, since the latter always assumes that there is a single source connecting all observable variables. In some sense one can think of Proposition 2 as the first level of the NPA hierarchy, but where the topology of the quantum network is explicitly taken into account.

In fact, consider a convex combination, where we with probability $v$ select a PR-box given by $p_{PR}(x_1,x_2 \vert s_1,s_2)=(1/2) \delta_{x_1\oplus x_2,s_1s_2}$ (assuming all variable dichotomic and assuming the values $0,1$), and with probability $1-v$ select the white noise distribution $p_{W}(x_1,x_2 \vert s_1,s_2)=1/4$. A numerical implementation of the SDP test in Proposition 2 shows that the resulting distribution is incompatible with the quantum network for any $v > 1/\sqrt{2}$, which corresponds exactly to the Tsirelson's bound of $2\sqrt{2}$ in the CHSH inequality. That is, the test in Proposition 2 is able to detect post-quantum correlations.

\subsection{\label{PosDefComplQuantum} Semidefinite completion of the observable covariance matrix}

Here we show that the observable covariance matrix $C_{\mathrm{observable}}$ always possesses a positive semidefinite completion $C_{\mathrm{completion}}$, by explicitly constructing one such completion. We first make a few observations.

As discussed in the main text, we define the observable covariance matrix as
\begin{equation}
\label{dfbsfssbf2}
\begin{split}
C_{\mathrm{observable}} := &   \sum_{m}\sum_{s_m}\mathrm{Cov}(Y^{(m,s_m)})  +\sum_{\substack{m,m' \\ m\neq m'}}\sum_{s_m,s'_{m'}} \mathrm{Cov}(Y^{(m,s_m)},Y^{(m',s'_{m'})}).
\end{split}
\end{equation}
With the conditional distribution defined by (\ref{QuantumWithInputsAgain}), it follows that the covariances $
\mathrm{Cov}(Y^{(m,s_m)})$ for each single child $c_m$ can be written
\begin{equation}
\label{hndghndg}
\begin{split}
\mathrm{Cov}(Y^{(m,s_m)})
= &   \sum_{x_{m,s_m}} Y^{(m,s_m)}_{x_{m,s_m}}{Y^{(m,s_m)}_{x_{m,s_m}}}^{\dagger} \Tr([A^{(m,s_m)}_{x_{m,s_m}}\otimes\hat{1}_{/c_m}]\rho)\\
&  - \sum_{x_{m,s_m}} \sum_{x'_{m,s_m}}  Y^{(m,s_m)}_{x_{m,s_m}}{Y^{(m,s_m)}_{x'_{m,s_m}}}^{\dagger}   \Tr([A^{(m,s_m)}_{x_{m,s_m}}\otimes\hat{1}_{/c_m}]\rho)    \Tr([A^{(m,s_m)}_{x'_{m,s_m}}\otimes\hat{1}_{/c_m}]\rho),
\end{split}
\end{equation}
where $\hat{1}_{/c_m}$ denotes the identity operator on $\mathcal{H}_{c_1}\otimes\cdots\otimes \mathcal{H}_{c_{m-1}}\otimes \mathcal{H}_{c_{m+1}}\otimes\cdots\otimes\mathcal{H}_{c_M}$.
For the `cross-covariances' between each pair of children $c_m,c_{m'}$ with  $m\neq m'$, we analogously get
\begin{equation}
\label{tuiktik}
\begin{split}
& \mathrm{Cov}(Y^{(m,s_m)},Y^{(m',s'_{m'})}) \\
= &   \sum_{x_{m,s_m}} \sum_{x'_{m',s'_{m'}}} Y^{(m,s_m)}_{x_{m,s_m}}  {Y^{(m',s'_{m'})}_{x'_{m',s'_{m'}}}}^{\dagger}\\
& \times \bigg[\Tr(A^{(m,s_m)}_{x_{m,s_m}}\otimes  A^{(m',s'_{m'})}_{x'_{m',s'_{m'}}} \otimes\hat{1}_{/c_mc_{m'}} \rho)- \Tr([A^{(m,s_m)}_{x_{m,s_m}}\otimes\hat{1}_{/c_m}]\rho)\Tr(  [A^{(m',s'_{m'})}_{x'_{m',s'_{m'}}} \otimes\hat{1}_{/c_m}]\rho)\bigg].
\end{split}
\end{equation}
It is convenient to define the operator
\begin{equation}
\label{DefQ}
\begin{split}
Q := \sum_{m}\sum_{s_m}\sum_{x_{m,s_m}} Y^{(m,s_m)}_{x_{m,s_m}}\otimes A^{(m,s_m)}_{x_{m,s_m}}\otimes\hat{1}_{/c_m}.
\end{split}
\end{equation}
 We moreover define
\begin{equation}
\label{qerveaqr}
\begin{split}
 R := &   \sum_{m}\sum_{s_m}\sum_{x_{m,s_m}}  Y^{(m,s_m)}_{x_{m,s_m}}{Y^{(m,s_m)}_{x_{m,s_m}}}^{\dagger} \Tr( [A^{(m,s_m)}_{x_{m,s_m}}\otimes\hat{1}_{/c_m}]\rho) \\
& -  \sum_{m}\sum_{s_m}\sum_{x_{m,s_m}}\sum_{x'_{m,s_m}}  Y^{(m,s_m)}_{x'_{m,s_m}}{Y^{(m,s_m)}_{x_{m,s_m}}}^{\dagger} \Tr( [A^{(m,s_m)}_{x'_{m,s_m}} A^{(m,s_m)}_{x_{m,s_m}}\otimes\hat{1}_{/c_m}]\rho).
 \end{split}
\end{equation}
The  completion, of the observable covariance matrix stemming from a given state and given collections of POVMs, can be stated as follows. 
\begin{Lemma}
\label{LePosDef}
Let $\rho$ be any density operator on $\mathcal{H} = \mathcal{H}_{c_1}\otimes\cdots\otimes\mathcal{H}_{c_M}$. Let $Q$ be defined as in (\ref{DefQ}), and $R$ as defined in (\ref{qerveaqr}), and $C_{\mathrm{observable}}$ as defined by  (\ref{dfbsfssbf2}), (\ref{hndghndg}) and (\ref{tuiktik}). Then it is the case that 
\begin{equation}
\label{fgbsfgb}
\Tr_{\mathcal{H}}\bigg[ \Big(Q -\Tr_{\mathcal{H}}(Q\rho)\Big) \Big(Q -\Tr_{\mathcal{H}}(Q\rho)\Big)^{\dagger} \rho \bigg]  + R =   C_{\mathrm{observable}} + C_{\mathrm{completion}},
\end{equation}
where
\begin{equation}
\label{DefCompletionAgain}
\begin{split}
C_{\mathrm{completion}} := & \sum_{m}\sum_{\substack{s_m,s'_{m} \\ s_m\neq s'_m}}\sum_{x_{m,s_m}} \sum_{x'_{m,s'_{m}}}  Y^{(m,s_m)}_{x_{m,s_m}} {Y^{(m,s'_{m})}_{x'_{m,s'_{m}}}}^{\dagger}\\
&\times\bigg[\Tr\Big( [A^{(m,s_m)}_{x_{m,s_m}}  A^{(m,s'_{m})}_{x'_{m,s'_{m}}}\otimes\hat{1}_{/c_m}] \rho\Big) -\Tr( [A^{(m,s_m)}_{x_{m,s_m}}\otimes\hat{1}_{/c_m}]\rho)\Tr( [A^{(m,s'_{m})}_{x'_{m,s'_{m}}}\otimes\hat{1}_{/c_m}]\rho)\bigg].
\end{split}
\end{equation}
Moreover
\begin{equation}
\label{PosDefCompl}
\begin{split}
& C_{\mathrm{observable}}+ C_{\mathrm{completion}}\geq 0,\quad R\geq 0,
\end{split}
\end{equation}
\begin{equation}
\label{gfnsfgnsf2}
\begin{split}
& P_{m,s_m}C_{\mathrm{completion}}P_{m,s_m} = 0,\quad m = 1,\ldots, M,\quad s_m = 1,\ldots,S_m,\\
& P_{m,s_m}C_{\mathrm{completion}}P_{m',s'_{m'}} = 0,\quad m,m' = 1,\ldots, M,\quad m\neq m',\quad s_m = 1,\ldots,S_m,\quad s'_{m'} = 1,\ldots, S_{m'},\\
\end{split}
\end{equation}
\begin{equation}
\label{tezuen2}
 C_{\mathrm{completion}}^{\dagger} = C_{\mathrm{completion}}.
\end{equation}
\end{Lemma}

\emph{Remark:} One may note that this lemma holds for any density operator $\rho$, i.e., we are not restricted to density operators on the form $\rho =\rho_{p_1}\otimes\cdots\otimes\rho_{p_N}$, which however is needed in the setting of Proposition 2.

\begin{proof}
We first use (\ref{DefQ}) to see that 
\begin{equation*}
\begin{split}
& \Tr_{\mathcal{H}}\bigg[ \Big(Q -\Tr_{\mathcal{H}}(Q\rho)\Big) \Big(Q -\Tr_{\mathcal{H}}(Q\rho)\Big)^{\dagger} \rho \bigg] \\
= & \Tr_{\mathcal{H}}(QQ^{\dagger}\rho) - \Tr_{\mathcal{H}}(Q\rho)\Tr_{\mathcal{H}}(Q\rho)^{\dagger}\\
=  & \sum_{m,m'}\sum_{s_m,s'_{m'}}\sum_{x_{m,s_m}} \sum_{x'_{m',s'_{m'}}}  Y^{(m,s_m)}_{x_{m,s_m}} {Y^{(m',s'_{m'})}_{x'_{m',s'_{m'}}}}^{\dagger}\\
& \quad\quad\times\bigg[\Tr\Big( [A^{(m,s_m)}_{x_{m,s_m}}  A^{(m',s'_{m'})}_{x'_{m',s'_{m'}}}\otimes\hat{1}_{/c_mc_{m'}}] \rho\Big) -\Tr( [A^{(m,s_m)}_{x_{m,s_m}}\otimes\hat{1}_{/c_{m}}]\rho)\Tr( [A^{(m',s'_{m'})}_{x'_{m',s'_{m'}}}\otimes\hat{1}_{/c_{m'}}]\rho)\bigg]\\
\end{split}
\end{equation*}
\begin{equation*}
\begin{split}
= & \sum_{m}\sum_{s_m}\sum_{x_{m,s_m}} \sum_{x'_{m,s_{m}}}  Y^{(m,s_m)}_{x_{m,s_m}} {Y^{(m,s_{m})}_{x'_{m,s_{m}}}}^{\dagger}\\
& \quad\quad\times\bigg[\Tr\Big( A^{(m,s_m)}_{x_{m,s_m}}  A^{(m,s_{m})}_{x'_{m,s_{m}}} \rho\Big) -\Tr( A^{(m,s_m)}_{x_{m,s_m}}\rho)\Tr( A^{(m,s_{m})}_{x'_{m,s_{m}}}\rho)\bigg]\\
& + \sum_{m}\sum_{\substack{s_m,s'_{m} \\ s_m\neq s'_m}}\sum_{x_{m,s_m}} \sum_{x'_{m,s'_{m}}}  Y^{(m,s_m)}_{x_{m,s_m}} {Y^{(m,s'_{m})}_{x'_{m,s'_{m}}}}^{\dagger}\\
& \quad\quad\times\bigg[\Tr\Big( A^{(m,s_m)}_{x_{m,s_m}}  A^{(m,s'_{m})}_{x'_{m,s'_{m}}} \rho\Big) -\Tr( A^{(m,s_m)}_{x_{m,s_m}}\rho)\Tr( A^{(m,s'_{m})}_{x'_{m,s'_{m}}}\rho)\bigg]\\
& +\sum_{\substack{m,m' \\ m\neq m'}}\sum_{s_m,s'_{m'}}\sum_{x_{m,s_m}} \sum_{x'_{m',s'_{m'}}}  Y^{(m,s_m)}_{x_{m,s_m}} {Y^{(m',s'_{m'})}_{x'_{m',s'_{m'}}}}^{\dagger}\\
& \quad\quad \times\bigg[\Tr\Big( A^{(m,s_m)}_{x_{m,s_m}}  A^{(m',s'_{m'})}_{x'_{m',s'_{m'}}} \rho\Big) -\Tr( A^{(m,s_m)}_{x_{m,s_m}}\rho)\Tr( A^{(m',s'_{m'})}_{x'_{m',s'_{m'}}}\rho)\bigg]\\
\end{split}
\end{equation*}
\begin{equation*}
\begin{split}
= & \sum_{m}\sum_{s_m}\sum_{x_{m,s_m}} \sum_{x'_{m,s_{m}}}  Y^{(m,s_m)}_{x_{m,s_m}} {Y^{(m,s_{m})}_{x'_{m,s_{m}}}}^{\dagger}\\
& \quad\quad\times\bigg[\Tr\Big( [A^{(m,s_m)}_{x_{m,s_m}}  A^{(m,s_{m})}_{x'_{m,s_{m}}}\otimes\hat{1}_{/c_m}] \rho\Big) -\Tr( [A^{(m,s_m)}_{x_{m,s_m}}\otimes\hat{1}_{/c_m}]\rho)\Tr( [A^{(m,s_{m})}_{x'_{m,s_{m}}}\otimes\hat{1}_{/c_m}]\rho)\bigg]\\
& + C_{\mathrm{completion}}\\
& +\sum_{\substack{m,m'\\ m\neq m'}}\sum_{s_m,s'_{m'}}\mathrm{Cov}(Y^{(m,s_m)},Y^{(m',s'_{m'})})
\end{split}
\end{equation*}

Hence,
\begin{equation*}
\begin{split}
& \Tr_{\mathcal{H}}\bigg[ \Big(Q -\Tr_{\mathcal{H}}(Q\rho)\Big) \Big(Q -\Tr_{\mathcal{H}}(Q\rho)\Big)^{\dagger} \rho \bigg] + R \\
& = 
  C_{\mathrm{completion}}\\
& +\sum_{\substack{m,m' \\ m\neq m'}}\sum_{s_m,s'_{m'}}\mathrm{Cov}(Y^{(m,s_m)},Y^{(m',s'_{m'})})\\
& +    \sum_{m}\sum_{s_m}\mathrm{Cov}(Y^{(m,s_m)}) \\
& = 
  C_{\mathrm{completion}} + C_{\mathrm{observable}}.
\end{split}
\end{equation*}
Hence, we have confirmed (\ref{fgbsfgb}).

Next, we wish to show that  $R\geq 0$  and $C_{\mathrm{completion}} + C_{\mathrm{observable}}\geq 0$. 
To this end, define
\begin{equation}
\label{DefR}
\begin{split}
\mathcal{R} :=&   \sum_{m}\sum_{s_m}\sum_{x_{m,s_m}} \\
& \Big[Y^{(m,s_m)}_{x_{m,s_m}} \otimes\hat{1}-  \sum_{x'_{m,s_m}}Y^{(m,s_m)}_{x'_{m,s_m}}\otimes A^{(m,s_m)}_{x'_{m,s_m}}  \Big] A^{(m,s_m)}_{x_{m,s_m}}  \Big[Y^{(m,s_m)}_{x_{m,s_m}} \otimes\hat{1}-  \sum_{x''_{m,s_m}}Y^{(m,s_m)}_{x''_{m,s_m}}\otimes A^{(m,s_m)}_{x''_{m,s_m}}  \Big]^{\dagger}.
 \end{split}
\end{equation}
One can confirm that  $R = \Tr_{\mathcal{H}}(\mathcal{R}\rho)$. Moreover, by the manner in which $\mathcal{R}$ is constructed in (\ref{DefR}), one can see that $\mathcal{R}\geq 0$. Consequently, $R = \Tr_{\mathcal{H}}(\mathcal{R}\rho)\geq 0$. One can also realize that 
\begin{equation*}
\begin{split}
& \Tr_{\mathcal{H}}\bigg[ \Big(Q -\Tr_{\mathcal{H}}(Q\rho)\Big) \Big(Q -\Tr_{\mathcal{H}}(Q\rho)\Big)^{\dagger} \rho \bigg]\geq 0.
\end{split}
\end{equation*}
Hence, the left hand side of (\ref{fgbsfgb}) is positive semidefinite, and thus is also the right hand side. We can thus conclude that $C_{\mathrm{completion}} + C_{\mathrm{observable}}\geq 0$. 

Next, we wish to prove (\ref{gfnsfgnsf2}).
First note that 
\begin{equation*}
P_{m'',s''_{m''}}Y^{(m,s_{m})}_{x_{m,s_{m}}} = \delta_{m'',m}\delta_{s''_{m},s_m}Y^{(m,s_{m})}_{x_{m,s_{m}}}.  
\end{equation*}
Hence
\begin{equation*}
\begin{split}
& P_{m'',s''_{m''}}C_{\mathrm{completion}}P_{m'',s''_{m''}}\\
 & =  \sum_{m}\sum_{\substack{s_m,s'_{m}\\ s_m\neq s'_m}}\sum_{x_{m,s_m}} \sum_{x'_{m,s'_{m}}} P_{m'',s''_{m''}}Y^{(m,s_m)}_{x_{m,s_m}} {Y^{(m,s'_{m})}_{x'_{m,s'_{m}}}}^{\dagger}P_{m'',s''_{m''}}\\
&\quad\times\bigg[\Tr\Big( A^{(m,s_m)}_{x_{m,s_m}}  A^{(m,s'_{m})}_{x'_{m,s'_{m}}} \rho\Big) -\Tr( A^{(m,s_m)}_{x_{m,s_m}}\rho)\Tr( A^{(m,s'_{m})}_{x'_{m,s'_{m}}}\rho)\bigg]\\
& =  0.
\end{split}
\end{equation*}
Similarly, for $m''\neq m'''$, it is the case that 
\begin{equation*}
\begin{split}
& P_{m'',s''_{m''}}C_{\mathrm{completion}}P_{m''',s'''_{m'''}}\\
&  =  \sum_{m}\sum_{\substack{s_m,s'_{m}\\ s_m\neq s'_m}}\sum_{x_{m,s_m}} \sum_{x'_{m,s'_{m}}} P_{m'',s''_{m''}} Y^{(m,s_m)}_{x_{m,s_m}} {Y^{(m,s'_{m})}_{x'_{m,s'_{m}}}}^{\dagger}P_{m''',s'''_{m'''}}\\
&\quad\times\bigg[\Tr\Big( A^{(m,s_m)}_{x_{m,s_m}}  A^{(m,s'_{m})}_{x'_{m,s'_{m}}} \rho\Big) -\Tr( A^{(m,s_m)}_{x_{m,s_m}}\rho)\Tr( A^{(m,s'_{m})}_{x'_{m,s'_{m}}}\rho)\bigg]\\
& =  0.
\end{split}
\end{equation*}
Hence, we have shown (\ref{gfnsfgnsf2}).

Finally, for the proof of (\ref{tezuen2}), one should keep in mind that $\Tr\big( A^{(m,s_m)}_{x_{m,s_m}}  A^{(m,s'_{m})}_{x'_{m,s'_{m}}} \rho\big)$ in general can be complex, and that $\Tr\big( A^{(m,s_m)}_{x_{m,s_m}}  A^{(m,s'_{m})}_{x'_{m,s'_{m}}} \rho\big)^{*} = \Tr\big(  A^{(m,s'_{m})}_{x'_{m,s'_{m}}}A^{(m,s_m)}_{x_{m,s_m}}\rho\big)$.
\begin{equation*}
\begin{split}
C_{\mathrm{completion}}^{\dagger} = & \sum_{m}\sum_{\substack{s_m,s'_{m}\\ s_m\neq s'_m}}\sum_{x_{m,s_m}} \sum_{x'_{m,s'_{m}}}  Y^{(m,s'_{m})}_{x'_{m,s'_{m}}} {Y^{(m,s_m)}_{x_{m,s_m}}}^{\dagger}\\
&\quad\times\bigg[\Tr\Big( A^{(m,s_m)}_{x_{m,s_m}}  A^{(m,s'_{m})}_{x'_{m,s'_{m}}} \rho\Big)^{*} -\Tr( A^{(m,s_m)}_{x_{m,s_m}}\rho)\Tr( A^{(m,s'_{m})}_{x'_{m,s'_{m}}}\rho)\bigg]\\
= & \sum_{m}\sum_{\substack{s_m,s'_{m}\\ s_m\neq s'_m}}\sum_{x_{m,s_m}} \sum_{x'_{m,s'_{m}}}  Y^{(m,s'_{m})}_{x'_{m,s'_{m}}} {Y^{(m,s_m)}_{x_{m,s_m}}}^{\dagger}\\
&\quad\times\bigg[\Tr\Big(  A^{(m,s'_{m})}_{x'_{m,s'_{m}}} A^{(m,s_m)}_{x_{m,s_m}}\rho\Big) -\Tr( A^{(m,s_m)}_{x_{m,s_m}}\rho)\Tr( A^{(m,s'_{m})}_{x'_{m,s'_{m}}}\rho)\bigg]\\
= & C_{\mathrm{completion}},
\end{split}
\end{equation*}
which demonstrates (\ref{tezuen2}).
\end{proof}

%%%%%%%%%%%%%%%%%%%%%%%%%%%%%%%%%%%%%%%%%%%%%%%%%

\subsection{Unobservable covariances}

As briefly mentioned in the main text, we regard the covariances $\mathrm{Cov}(Y^{(m,s_m)},Y^{(m,s^{\prime}_{m})})$  for $s_{m}\neq s^{\prime}_{m}$ as being `unobservable'. The basic rationale is that each child generally only can measure one single POVM at a time, and it thus seems difficult to give a direct observational meaning to covariances between two measurements that exclude each other. For the classical case, we did in Section \ref{SecProofIdea} observe that we in some sense can overcome this by considering extensions of the underlying random variables. However, as discussed above, it is difficult to see how this strategy could be transferred to the quantum setting. One can certainly consider special cases where simultaneous measurements are feasible, such as two POVMs corresponding to the eigenprojectors of two commuting observables. However, in more general cases, such as eigenprojectors of non-commutative observables, it seems challenging to provide a direct interpretation of $\mathrm{Cov}(Y^{(m,s_m)},Y^{(m,s^{\prime}_{m})})$ in terms of standard (strong) measurements.

 Similar remarks apply to the explicit construction of the completion $C_{\mathrm{completion}}$  in (\ref{DefCompletionAgain}).
While $C_{\mathrm{observable}}$ in (\ref{dfbsfssbf2}) has a direct physical interpretation in terms of (strong) measurements  of the POVMs  $A^{(m,s_m)}$, it seems problematic to provide $C_{\mathrm{completion}}$ with such an interpretation, due to the terms $\Tr\big( A^{(m,s_m)}_{x_{m,s_m}}  A^{(m,s^{\prime}_{m})}_{x^{\prime}_{m,s^{\prime}_{m}}} \rho\big)$ in (\ref{DefCompletionAgain}). One may for example note that the products $Q := A^{(m,s_m)}_{x_{m,s_m}}  A^{(m,s^{\prime}_{m})}_{x^{\prime}_{m,s^{\prime}_{m}}}$ generally cannot be interpreted as POVM elements (since they typically would not be positive semidefinite), and not even as observables (since they generally would not be Hermitian). One could in principle imagine schemes where one indirectly determines $\Tr(Q\rho)$ via expectation values of the observables $Q+Q^{\dagger}$ and $i(Q-Q^{\dagger})$, or via weak measurements. However, here we will not attempt to make conceptual sense,  within the current context,  of such alternative constructions. For our purpose it suffices that $C_{\mathrm{completion}}$, on a purely mathematical level, completes $C_{\mathrm{observable}}$ into a positive semidefinite matrix.

%%%%%%%%%%%%%%%%%%%%%%%%%%%%%%%%%%%%%%%%%%%%%%%%%

\subsection{Semidefinite decompositions for quantum networks with inputs}

For convenience we here restate Proposition 2 contained in the main text.

\begin{Proposition}
\label{PropQuantumBipartiteWithChoiceAgain}
Let the conditional distribution $P(x_1,\ldots,x_M|s_1,\ldots,s_M)$ be compatible, in the sense of (\ref{QuantumWithInputsAgain}), with  the quantum causal structure $G$ with parents $p_1,\ldots,p_N$ and children $c_1,\ldots,c_M$, with associated inputs $s_1,\ldots,s_M$. Assume that each child $c_m$, and each input $s_m$, is assigned a feature map $Y^{(m,s_m)}$ into a vector space $\mathcal{V}_{m,s_m}$. Let the operator $C_{\mathrm{observable}}$  on $\mathcal{V} := \bigoplus_{m=1}^{M}\bigoplus_{s_m}\mathcal{V}_{m,s_m}$ be as defined in (\ref{dfbsfssbf2}).
Then there exist operators  $C_{\mathrm{completion}}$, $R$ and $(C_n)_{n=1}^{N}$ on $\mathcal{V}$, such that $C_{\mathrm{completion}}$ satisfies  (\ref{gfnsfgnsf2}) and
\begin{equation}
\label{ConditionCompletionAgain}
C: = C_{\mathrm{observable}} + C_{\mathrm{completion}}\geq 0,
\end{equation}
and
\begin{eqnarray}
\label{sfgbsfgb}
C & = & R + \sum_{n=1}^{N}C_n,\quad R\geq 0,\quad C_n\geq 0,\\
\label{fgndnfhdquantum}
& & P^{(n)}C_n P^{(n)} = C_n,\quad R = \sum_{m,s_m}P_{m,s_m}R P_{m,s_m},\\
\label{mhdghmggd}
& & P^{(n)}:= \sum_{m\in\mathcal{C}_n}\sum_{s_m=1}^{S_m}P_{m,s_m},
\end{eqnarray}
where $P_{m,s_m}$ is the projector onto $\mathcal{V}_{m,s_m}$, and where $\mathcal{C}_n$ denotes the children having parent $p_n$ in the given DAG $G$.
\end{Proposition}

\emph{Remark:}  Analogous to  the completion  in the classical version in Supplementary Proposition \ref{PropClassicalWithInputs}, it follows from the requirement that $C$ is positive semidefinite, that $C_{\mathrm{completion}}$ is Hermitian. Thus, in any  search for a completion, one can assume that is Hermitian. As stated in Supplementary Lemma \ref{LePosDef}, the specific choice of completion  in (\ref{DefCompletionAgain}) is indeed also Hermitian.

\begin{proof}

By the assumption that the conditional distribution $P(x_1,\ldots,x_M|s_1,\ldots,s_M)$ is compatible with  the quantum causal structure $G$ with parents $p_1,\ldots,p_N$ and children $c_1,\ldots,c_M$, with associated inputs $s_1,\ldots,s_M$, there exists a state as in (\ref{DAGGstatesAgain}) and POVMs on the form (\ref{DAGGPOVMinputs}), such that $P(x_1,\ldots,x_M|s_1,\ldots,s_M)$ can be written as in (\ref{QuantumWithInputsAgain}).  With the assigned feature maps $Y^{m,s_m}_{x_{m,s_m}}$ to the outcomes $x_{m,s_m}$ corresponding to POVM $A^{(m,s_m)}_{x_{m,s_m}}$, the  observable covariance $C_{\mathrm{observable}}$ is defined as in (\ref{dfbsfssbf2}). By Supplementary Lemma \ref{LePosDef} we know that $C_{\mathrm{completion}}$, defined in (\ref{DefCompletionAgain}) is such that $C: = C_{\mathrm{observable}} + C_{\mathrm{completion}}\geq 0$ and thus (\ref{ConditionCompletionAgain}) holds. By Supplementary Lemma \ref{LePosDef} we moreover know that  $C_{\mathrm{completion}}$ satisfies (\ref{gfnsfgnsf2}).

Next, we consider $Q$, defined in (\ref{DefQ}), and note that 
\begin{equation}
\begin{split}
\Tr_{\mathcal{H}}\bigg[ \Big(Q -\Tr_{\mathcal{H}}(Q\rho)\Big) \Big(Q -\Tr_{\mathcal{H}}(Q\rho)\Big)^{\dagger} \rho \bigg]  = & \Tr_{\mathcal{H}}(QQ^{\dagger}\rho)- \Tr_{\mathcal{H}}(Q\rho)\Tr_{\mathcal{H}}(Q^{\dagger}\rho)\\
= &\sum_{n=1}^{N}C_n,
\end{split}
\end{equation}
where the last equality follows by  Supplementary Lemma  \ref{nfkelnlfk}, where $C_n$  are defined as in (\ref{nfdlkbnfd}).

By equation (\ref{fgbsfgb}) in Supplementary Lemma  \ref{LePosDef}, we can conclude that $C_{\mathrm{observable}} + C_{\mathrm{completion}} = C = R + \sum_{n=1}^{N}C_n$, with $R$, defined by (\ref{qerveaqr}). 
 Furthermore, Supplementary Lemma  \ref{LePosDef} yields that $R$ is positive semidefinite, and Supplementary Lemma \ref{nfkelnlfk} yields that each $C_n$ is positive semidefinite. Hence we can conclude that (\ref{sfgbsfgb}) holds.

Next we shall show that  $P^{(n)}C_nP^{(n)} =C_n$, for the projectors $P^{(n)}$ defined  in (\ref{mhdghmggd}) with respect to the given DAG $G$.
For $2\leq n\leq N-1$, recall the definition of $C_n$ in (\ref{nfdlkbnfd}). With the definition of $Q$ in (\ref{DefQ}), it follows that 
\begin{equation}
\label{zmrzum}
\begin{split}
  C_n := & \sum_{m,m'}\sum_{s_m,s'_{m'}}\sum_{x_{m,s_m}} \sum_{x'_{m',s'_{m'}}}  Y^{(m,s_m)}_{x_{m,s_m}}  {Y^{(m',s'_{m'})}_{x'_{m',s'_{m'}}}}^{\dagger} \Tr_{p_n,\ldots,p_N}\bigg[ W^{m,s_m,n}_{x_{m,s_m}} {W^{m',s'_{m'},n}_{x'_{m',s'_{m'}}}}^{\dagger}  [\rho_{p_n}\otimes\cdots\otimes \rho_{p_N}]\bigg],\\
\end{split}
\end{equation}
where
\begin{equation}
\begin{split}
 W^{m,s_m,n}_{x_{m,s_m}}& :=  \Tr_{p_1,\ldots, p_{n-1}}([A^{(m,s_m)}_{x_{m,s_m}}\otimes\hat{1}_{/c_m}][\rho_{p_1}\otimes\cdots \otimes\rho_{p_{n-1}}]) \\
& - \hat{1}_{p_n}\otimes\Tr_{p_1,\ldots,p_n}([A^{(m,s_m)}_{x_{m,s_m}}\otimes\hat{1}_{/c_m}][\rho_{p_1}\otimes\cdots \otimes\rho_{p_n}]).
\end{split}
\end{equation}
Suppose that $m\notin \mathcal{C}_n$. This implies that $\hat{1}_{p_n}\otimes \Tr_{p_n}([A_{x_{m,s_m}}^{(m,s_m)}\otimes\hat{1}_{/c_m}]\rho_{p_n})= A_{x_{m,s_m}}^{(m,s_m)}\otimes\hat{1}_{/c_m}$.
Consequently $W^{m,s_m,n}_{x_{m,s_m}}  = 0$, for all $s_m$  and all $x_{m,s_m}$.  Hence, if $m\notin \mathcal{C}_n$ or if $m'\notin \mathcal{C}_n$, then
\begin{equation}
\label{adsvsd}
W^{m,s_m,n}_{x_{m,s_m}}{W^{m',s'_{m'},n}_{x'_{m',s'_{m'}}}}^{\dagger} = 0,\quad \forall s_m,s'_{m'}, x_{m,s_m},x'_{m',s'_{m'}}.
\end{equation}
Recall that $Y^{(m,s_m)}_{x_{m,s_m}}$ is supported on $\mathcal{V}_{m,s_m}$. By comparing (\ref{zmrzum}) with with (\ref{adsvsd}) one can realize that $P_{m.s_m} C_n P_{m',s'_{m'}} = 0$ if $m\notin \mathcal{C}_n$ or if $m'\notin \mathcal{C}_n$, for all $s_m$ and $s'_{m'}$. Hence
\begin{equation*}
\begin{split}
C_n = & \sum_{m=1}^{M}\sum_{s_m=1}^{S_m} P_{m,s_m} C_n \sum_{m'= 1}^{M}\sum_{s'_{m'}= 1}^{S_{m'}}P_{m',s'_{m'}}\\
 = & \sum_{m\in \mathcal{C}_n}\sum_{s_m= 1}^{S_m} P_{m,s_m} C_n\sum_{m'\in\mathcal{C}_n}\sum_{s'_{m'}=1}^{S_m} P_{m',s'_{m'}}\\
= & P^{(n)}C_n P^{(n)}.
\end{split}
\end{equation*} 
We can thus  conclude that $P^{(n)}C_nP^{(n)} = C_n$, for $2\leq n\leq N-1$. 
The proofs for the cases $n = 1$ and $n = N$ are analogous.

As the final step we need to  show that $R$, defined in (\ref{qerveaqr}), satisfies the decomposition $R = \sum_{m,s_m}P_{m,s_m}RP_{m,s_m}$. Since $Y^{(m,s_m)}_{x_{m,s_m}}$ is supported on $\mathcal{V}_{m,s_m}$, it follows that $P_{m',s'_{m'}}Y^{(m,s_m)}_{x_{m,s_m}} = \delta_{m,m'}\delta_{s_m,s'_m}Y^{(m,s_m)}_{x_{m,s_m}}$ for all $x_{m,s_m}$. By comparison with (\ref{qerveaqr}) it follows that $P_{m,s_m}RP_{m',s'_{m'}} = \delta_{m,m'}\delta_{s_m,s'_m}P_{m,s_m}RP_{m,s_m}$, and thus $R = \sum_{m,s_m}P_{m,s_m}RP_{m,s_m}$.

\end{proof}

\subsection{Semidefinite test for quantum networks with inputs}

Analogous to how the case of quantum networks without inputs in Proposition 1 gives rise to a semidefinite test, Supplementary Proposition \ref{PropQuantumBipartiteWithChoiceAgain} can also be turned into a semidefinite test that allows us to rule out hypothetical  causal structures as explanations of a given conditional distribution $P$. Supplementary Proposition \ref{PropQuantumBipartiteWithChoiceAgain} can be phrased schematically as follows: If $P$ iscompatible with $G$, then for every choice of feature maps there exist $C_{\textrm{completion}}$, $R$, and $(C_n)_n$ such that the `conditions' [(\ref{ConditionCompletionAgain}) - (\ref{mhdghmggd})]
are satisfied. 
This can equivalently be rephrased as follows:  If there exists a choice of feature maps for which the  `conditions' fail for all choices of $C_{\textrm{completion}}$, $R$, and $(C_n)_n$, then $P$ is not compatible with $G$. This alternative formulation is the basis of the semidefinite test.
It is worth stressing that one has to test \emph{all} completions $C_{\textrm{completion}}$ before a given quantum causal structure $G$ can be excluded as an explanation of an observed conditional distribution $P$.  
In particular, when applying the semidefinite test, one cannot in general fix the completion  $C_{\mathrm{completion}}$ to be (\ref{DefCompletionAgain}). For example, if the conditional distribution is obtained via some quantum state and families of POVMs (not  necessarily compatible with $G$), then one should not be tempted to use the completion in (\ref{DefCompletionAgain}) for the semidefinite test, since there may exist some \emph{other} states and collections of POVMs that would be compatible with $G$, and yield the same distribution.

By comparison of Supplementary Proposition \ref{PropQuantumBipartiteWithChoiceAgain} and Supplementary Proposition \ref{PropClassicalWithInputs}, one can see that the conditions of the completion and semidefinite decomposition is identical in both cases. We can thus conclude that even though we allow for measurement settings, the semidefnite test does not distinguish classical and quantum networks. This may seem surprising, since including measurement setting in many cases yield such distinctions. In view of this, it is worth pointing out that although the semidefinite test has identical structure in both the classical and quantum case, this does not exclude the possibility that the observable covariance matrix \emph{per se} may contain information that would allow for distinctions in some cases. More precisely, it may be the case that the set of observable covariance matrices that are compatible with a quantum network is larger than the set compatible with the corresponding classical network, even though the semidefinite test is oblivious to such differences.

\section{Alternative tests for quantum networks with inputs}

The semidefinite test for inputs in Proposition 2 is in some sense proved independently of the semidefinite test without inputs in Proposition 1 (although both depend on Supplementary Lemma \ref{nfkelnlfk}).  
However, one can consider methods to directly use Proposition 1 in order to generate tests also for the case of inputs.
Loosely speaking, from a setting with inputs we can generate a family of settings without inputs, onto which we can apply Proposition 1. If any of these sub-tests would fail, then we have falsified the hypothetical DAG as an explanation of the observed covariances. Here we consider two such alternative tests (or rather one test, and a generalization of it) and we show that none of these two alternatives is stronger than the test stemming from Proposition 2.  

Since these  alternative tests all have the structure of a collection of separate tests, while Proposition 2 in some sense can be regarded as yielding a `collective' test, it seems reasonable to expect that Proposition 2 yields strictly stronger tests. Although an interesting question, we leave this for future investigations, and focus on showing that Proposition 2 is not weaker than these alternatives.

\subsection{Tests from local selections}

Suppose that we have a setup with choices of inputs. Consider the case that each child $c_m$ fixes a specific input $s_m$. 
By fixing the inputs we effectively obtain a setting without inputs, onto which we can apply the test of Proposition  1. 
In other words, if there would exist a collective choice of inputs $\overline{s} = (s_1,\ldots, s_M)$ for which the semidefinite test of Proposition 1 would fail, then it means that the hypothetical DAG has been falsified. Here we show that this alternative `selection-test' is not stronger than that of Proposition 2.
We shall do this by showing that if the `full' observable covariance matrix satisfies the semidefinite decomposition in the sense of Proposition 2, then the selection test is also satisfied. Equivalently: if the selection test fails, then the test of Proposition 2 will also fail. Hence, the selection test is not stronger than the `full' test.  

For a given DAG $G$ and collection of local POVMs $\{A^{(m,s_m)}_{x_m}\}_{x_m}$ we have the conditional distribution of outcomes
\begin{equation}
\label{neaunbea}
P(x_1,\ldots,x_M|s_1,\ldots,s_M) :=  \Tr([A^{(1,s_1)}_{x_1}\otimes\cdots\otimes A^{(M,s_M)}_{x_M}]\rho),\quad\quad \rho := \rho_{p_1}\otimes\cdots\otimes\rho_{p_N}.
\end{equation}
Let us also assume a collection of feature maps $Y^{(m,s_m)}$.
For each choice of measurement inputs $\overline{s} = (s_1,\ldots, s_M)$ we define the corresponding covariance matrix with respect to (\ref{neaunbea}) as
\begin{equation}
\label{fgdsrtnsgfrn}
\begin{split}
C^{(\overline{s})} := &   \sum_{m}\mathrm{Cov}(Y^{(m,s_m)})  +\sum_{\substack{m,m' \\ m\neq m'}} \mathrm{Cov}(Y^{(m,s_m)},Y^{(m',s_{m'})}),
\end{split}
\end{equation}
with $\mathrm{Cov}(Y^{(m,s_m)})$ as in (\ref{hndghndg}), and $\mathrm{Cov}(Y^{(m,s_m)},Y^{(m',s_{m'})})$ as in (\ref{tuiktik}) with $\rho := \rho_{p_1}\otimes\cdots\otimes\rho_{p_N}$.
Similarly, we define the `full' observable covariance matrix  with respect to (\ref{neaunbea}) as
\begin{equation}
\label{sdfbsfgbgf}
\begin{split}
C_{\mathrm{observable}} := &   \sum_{m}\sum_{s_m}\mathrm{Cov}(Y^{(m,s_m)})  +\sum_{\substack{m,m' \\ m\neq m'}}\sum_{s_m,s'_{m'}} \mathrm{Cov}(Y^{(m,s_m)},Y^{(m',s'_{m'})}).
\end{split}
\end{equation}
The following proposition essentially states that if (\ref{sdfbsfgbgf})  satisfies the semidefinite decomposition of Proposition 2, then (\ref{fgdsrtnsgfrn}) satisfies the semidefinite decomposition  in the sense of Proposition  1.
\begin{Proposition}
\label{PropSelection}
Let $G$ be a DAG, and assume an underlying conditional distribution (\ref{neaunbea}), as well as a collection of feature maps $Y^{(m,m_s)}$ into spaces $\mathcal{V}_{m,s_m}$.
If $C_{\mathrm{observable}}$ defined in (\ref{sdfbsfgbgf}) is such that there exist operators $C_{\mathrm{completion}}$, $R$, $(C_n)_{n=1}^{N}$ such that
\begin{equation}
\begin{split}
\label{ghndghdm}
& P_{m,s_m}C_{\mathrm{completion}}P_{m,s_m} = 0,\quad m = 1,\ldots, M,\quad s_m = 1,\ldots,S_m,\\
& P_{m,s_m}C_{\mathrm{completion}}P_{m',s'_{m'}} = 0,\quad m,m' = 1,\ldots, M,\quad m\neq m',\quad s_m = 1,\ldots,S_m,\quad s'_{m'} = 1,\ldots, S_{m'},\\
\end{split}
\end{equation}
and 
\begin{eqnarray}
& & C  =  C_{\mathrm{observable}} + C_{\mathrm{completion}},\\
\label{numuktmgutk} & & C  =  R + \sum_{n=1}^{N}C_n,\quad R\geq 0,\quad C_n\geq 0,\\
\label{fdvdfsbfsgn}
 & &  P^{(n)}C_n P^{(n)} = C_n,\\
\label{vfdbadba}
& &  R  =  \sum_{m,s_m}P_{m,s_m}R P_{m,s_m},
\end{eqnarray}
where $P^{(n)}  := \sum_{m\in\mathcal{C}_n}\sum_{s_m=1}^{S_m}P_{m,s_m}$ and $P_{m,s_m}$ is the projector onto $\mathcal{V}_{m,s_m}$, 
and $\mathcal{C}_n$ denotes the children with parent $p_n$ in $G$,
then for every $\overline{s} = (s_1,\ldots,s_M)$, there exist operators $R^{(\overline{s})}$ and $(C_n^{(\overline{s})})_{n=1}^{N}$, such that 
\begin{eqnarray}
\label{fdbflfgnknm} & & C^{(\overline{s})} =  R^{(\overline{s})}  + \sum_{n=1}^{N}C_n^{(\overline{s})},\quad R^{(\overline{s})} \geq 0,\quad C_n^{(\overline{s})}\geq 0\\
\label{gnsfgnsfgn}
& &  C_n^{(\overline{s})}=  P^{(n,\overline{s})} C_n^{(\overline{s})} P^{(n,\overline{s})},\\
\label{utigfitlfvugl}
& & R^{(\overline{s})}  =  \sum_{m}P^{(\overline{s})}_{m}R^{(\overline{s})}P^{(\overline{s})}_{m},
\end{eqnarray}
where $P^{(n,\overline{s})}  :=   \sum_{m\in\mathcal{C}_n}P^{(\overline{s})}_m$,  and $P^{(\overline{s})}_m :=  P_{m,s_m}$, and 
where  $C^{(\overline{s})}$ is as defined in (\ref{fgdsrtnsgfrn}).
\end{Proposition}
\emph{Remark:} We can equivalently say that if $C^{(\overline{s})}$ fails to satisfy the semidefinite decomposition  (\ref{fdbflfgnknm}), then $C_{\mathrm{observable}}$ fails to satisfy the semidefinite decomposition  (\ref{numuktmgutk}). In other words, the `selection-test' is not stronger than the test resulting from Proposition 2.

\begin{proof}
Define $P^{(\overline{s})} := \sum_{m=1}^{M}P^{(\overline{s})}_m = \sum_{m=1}^{M}P_{m,s_m}$. 
The first step of the proof is to show that $P^{(\overline{s})}CP^{(\overline{s})} = C^{(\overline{s})}$.
To this end we first observe that 
\begin{equation}
\label{fgnfgndnfmd}
\begin{split}
&  P^{(n,\overline{s})}P^{(\overline{s})}= P^{(\overline{s})} P^{(n,\overline{s})} = P^{(\overline{s})}P^{(n)} =  P^{(n)}P^{(\overline{s})}  = P^{(n,\overline{s})},\\
& P^{(\overline{s})}P_{m',s'_{m'}}  = P_{m',s'_{m'}} P^{(\overline{s})} = \delta_{s_{m'},s'_{m'}}P_{m',s'_{m'}}.
\end{split}
\end{equation}
Next we note that  by  (\ref{ghndghdm}) it follows that $P^{(\overline{s})}C_{\mathrm{completion}}P^{(\overline{s})}  =   0$, which  in turn yields
 \begin{equation}
 \label{fnsfgnsfgn}
 P^{(\overline{s})}CP^{(\overline{s})}   =  P^{(\overline{s})} C_{\mathrm{observable}}P^{(\overline{s})} 
 =    \sum_{m = 1}^{M}P^{(\overline{s})}_{m}  C_{\mathrm{observable}}P^{(\overline{s})}_{m}  +    \sum_{m,m':m\neq m'} P^{(\overline{s})}_{m}  C_{\mathrm{observable}}P^{(\overline{s})}_{m'}  =  C^{(\overline{s})}.
\end{equation}
We can thus conclude that $P^{(\overline{s})} C P^{(\overline{s})}  = C^{(\overline{s})}$.

Next we will show that the decomposition (\ref{numuktmgutk}) leads to the decompsiton (\ref{fdbflfgnknm}) of $C^{(\overline{s})}$.
For this purpose we define $R^{(\overline{s})}  :=  P^{(\overline{s})}RP^{(\overline{s})}$ and $C_n^{(\overline{s})}  :=  P^{(\overline{s})}C_n P^{(\overline{s})}$.
We first observe 
$C^{(\overline{s})} =  P^{(\overline{s})}CP^{(\overline{s})} =  P^{(\overline{s})}RP^{(\overline{s})} + \sum_{n=1}^{N}P^{(\overline{s})}C_n P^{(\overline{s})}=  R^{(\overline{s})}  + \sum_{n=1}^{N}C_n^{(\overline{s})}$, 
where the second equality follows by (\ref{numuktmgutk}).
Next we wish to show that the operators $R^{(\overline{s})}$ and  $C_n^{(\overline{s})}$ satisfy all the required conditions. 
From (\ref{numuktmgutk}) we know that $R\geq 0$ and $C_n\geq 0$. By comparing with the definition of $R^{(\overline{s})}$ and $C_n^{(\overline{s})} $ we can conclude that $R^{(\overline{s})} \geq 0$ and $C_n^{(\overline{s})} \geq 0$.
Next we wish to show that  $R^{(\overline{s})}$ satisfies  property (\ref{utigfitlfvugl}). From  (\ref{vfdbadba})  we know that
$R = \sum_{m',s'_{m'}}P_{m',s'_{m'}}R P_{m',s'_{m'}}$. From $P^{(\overline{s})}P_{m',s'_{m'}} = \delta_{s_{m'},s'_{m'}}P_{m',s'_{m'}}$, it thus follows that  
$R^{(\overline{s})}  =  P^{(\overline{s})}RP^{(\overline{s})}
=  \sum_{m}P^{(\overline{s})}_{m}R^{(\overline{s})}P^{(\overline{s})}_{m}$.
Hence (\ref{utigfitlfvugl}) is satisfied.
Finally, we should show that $C_n^{(\overline{s})}$ satisfies (\ref{gnsfgnsfgn}).
By (\ref{fdvdfsbfsgn}) we know that  $P^{(n)}C_n P^{(n)} =  C_n$. By combining this with the observations in (\ref{fgnfgndnfmd}) we obtain $C_n^{(\overline{s})}  =  P^{(n,\overline{s})}  C_n^{(\overline{s})} P^{(n,\overline{s})}$, and thus   (\ref{gnsfgnsfgn}) is satisfied.
\end{proof}

\subsection{Tests from local random selections}

In the previous section we created a family of tests without inputs by fixing a specific selection of inputs. Here we shall generalize this by consider a scheme where we select the inputs according to  a local probability distribution. The previous setting is obtained as a special case when the distributions deterministically single out one specific choice of POVM.

For this scenario we assume that all the possible POVMs of child $c_m$ have the same number of outcomes irrespective of the choice $s_m$. In other words, instead of 
$\{A^{(m,s_m)}_{x_{m,s_m}}\}_{x_{m,s_m}}$ we can write $\{A^{(m,s_m)}_{x_{m}}\}_{x_{m}}$. Hence, the range of the index $x_m$ does not depend on $s_m$.  One should note that this is not a restriction, since if the number of POVM elements differ,  one can `pad' the POVMs with zero-operators, such that they all have the same number of elements. 
The joint conditional distribution is
\begin{equation}
\label{dfbsfgnsfg}
P(x_1,\ldots,x_M|s_1,\ldots,s_M) :=  \Tr([A^{(1,s_1)}_{x_1}\otimes\cdots\otimes A^{(M,s_M)}_{x_M}]\rho),\quad\quad \rho := \rho_{p_1}\otimes\cdots\otimes\rho_{p_N}.
\end{equation}
We let the local  input $s_m$ be selected according to the distribution $(q^{(m)}_{s_m})_{s_m}$. The resulting effective POVM   $\{\overline{A}^{(m)}_{x_{m}}\}_{x_{m}}$ is defined by 
\begin{equation}
\label{nhkjsfgngfsn}
\overline{A}^{(m)}_{x_{m}} := \sum_{s_m}q^{(m)}_{s_m} A^{(m,s_m)}_{x_m}.
\end{equation}
 Hence, for each choice of local distributions $q^{(1)},\ldots,q^{(M)}$ we have fixed local  POVMs  $\{\overline{A}^{(m)}_{x_{m}}\}_{x_m}$,  and can thus apply the  test of Proposition  2. Hence, if we can find one collection of local distribution $q^{(1)},\ldots,q^{(M)}$ that would violate this test, then we have falsified the given DAG $G$ as an explanation of the observed covariances. Compared to the previous section, we do in this case not only get a finite collection of semidefinite tests, but rather a continuum of them.
However, here we shall show that  this `local random test' still is not stronger the test of Proposition 2.
We follow the same strategy as in the previous section, and show that if the `full' test is satisfied, then the local random test is also  satisfied. Hence, if the local random test is falsified, then the full test is also falsified.

For the semidefinite test without inputs, let us choose feature maps $\overline{Y}^{(m)}$ into spaces $\overline{\mathcal{V}}_m$. We let $\overline{Y}:= \sum_{m=1}^{M}\overline{Y}^{(m)}$. With respect to the distribution 
\begin{equation}
\label{gqrdrnrnnr}
\overline{P}(x_1,\ldots,x_M) = \Tr([\overline{A}^{(1)}_{x_1}\otimes\cdots\otimes \overline{A}^{(M)}_{x_M}][\rho_{p_1}\otimes\cdots\otimes\rho_{p_N}]),
\end{equation}
we consider the covariance
\begin{equation}
\label{ndfjnbdlbjg}
\begin{split}
\mathrm{Cov}_{\overline{P}}(\overline{Y}) := & \sum_{m}\mathrm{Cov}_{\overline{P}}(\overline{Y}_m) +\sum_{m,m':m\neq m'}\mathrm{Cov}_{\overline{P}}(\overline{Y}_m,\overline{Y}_{m'})\\
%%%
= & \sum_m \sum_{x_m} \overline{Y}^{(m)}_{x_m} {\overline{Y}_{x_m}^{(m)}}^{\dagger}\Tr([\overline{A}^{(m)}_{x_m}\otimes \hat{1}_{/c_m} ]\rho) - \sum_m \sum_{x_m,x'_{m}} \overline{Y}^{(m)}_{x_m} {\overline{Y}_{x'_m}^{(m)}}^{\dagger}\Tr([\overline{A}^{(m)}_{x_m}\otimes \hat{1}_{/c_m} ]\rho)\Tr([\overline{A}^{(m)}_{x'_m}\otimes \hat{1}_{/c_m} ]\rho)\\
& +\sum_{m,m':m\neq m'}\sum_{x_m,x'_{m'}} \overline{Y}^{(m)}_{x_m} {\overline{Y}_{x'_{m'}}^{(m')}}^{\dagger}\bigg[\Tr([\overline{A}^{(m)}_{x_m}\otimes \overline{A}^{(m')}_{x'_{m'}}\otimes\hat{1}_{/c_mc_{m'}}]\rho)    -
\Tr([\overline{A}^{(m)}_{x_m}\otimes \hat{1}_{/c_m}]\rho)  \Tr([\overline{A}^{(m')}_{x'_{m'}}\otimes\hat{1}_{/c_{m'}}]\rho)  
\bigg],
\\
& \rho := \rho_1\otimes\cdots\otimes\rho_N,
\end{split}
\end{equation}
where we have added the subscript $\overline{P}$ to emphasize that the covariance is calculated with respect to the distribution (\ref{gqrdrnrnnr}). 
For the corresponding `full' test with inputs, we have to have one feature-space $\mathcal{V}_{m,s_m}$ for each child $c_m$ and each choice of measurement $s_m$. 
Since the number of outcomes of the local measurements are the same for all $s_m$, we let  $\dim\mathcal{V}_{m,s_m} = \dim\overline{\mathcal{V}}_m$.
In other words, we let all $\mathcal{V}_{m,s_m}$ be `copies' of $\overline{\mathcal{V}}_{m}$. In a similar manner we let the feature maps $Y^{(m,s_m)}$ in some sense be copies of $\overline{Y}^{(m)}$. The idea is that the different choices of $s_m$ should all give rise to the `same outcome' on the level of the covariance matrix. More precisely, we define mappings from $\overline{\mathcal{V}}_{m}$ to $\mathcal{V}_{m,s_m}$ to construct  $Y^{(m,s_m)}$ from $\overline{Y}^{(m)}$ via partial isometries $V_{m,s_m}$ defined in (\ref{gngdndg}).
\begin{Proposition}
Let $G$ be a DAG, and assume an underlying conditional distribution (\ref{neaunbea}). We assume a collection of feature maps $Y^{(m,s_m)}$ into spaces $\mathcal{V}_{m,s_m}$, and a collection of  feature maps $\overline{Y}_m$ into spaces $\overline{\mathcal{V}}_m$, where we assume linear mappings  $V_{m,s_m}:\bigoplus_{m=1}^{M}\overline{\mathcal{V}}_m \rightarrow \bigoplus_{m=1}^{M}\bigoplus_{s_m = 1}^{S_m}\mathcal{V}_{m,s_m}$, such that 
\begin{equation}
\label{gngdndg}
Y^{(m,s_m)}_{x_m} :=V_{m,s_m}\overline{Y}^{(m)}_{x_m},\quad\quad  V_{m,s_m}^{\dagger}V_{m,s_m} = \overline{P}_{m},\quad\quad   V_{m,s_m}V_{m,s_m}^{\dagger} = P_{m,s_m},
\end{equation}
where $\overline{P}_m$ is the projector onto $\overline{\mathcal{V}}_m$, and  $P_{m,s_m}$ is the projector onto $\mathcal{V}_{m,s_m}$. Let $\mathrm{Cov}_{\overline{P}}(\overline{Y})$ be as defined in (\ref{ndfjnbdlbjg}), with respect to (\ref{gqrdrnrnnr}) where $\{\overline{A}^{(m)}_{x_m}\}_{x_m}$ is as in (\ref{nhkjsfgngfsn}) for given distributions $q^{(1)},\ldots,q^{(M)}$.
If $C_{\mathrm{observable}}$, defined in (\ref{sdfbsfgbgf}), is such that there exist operators $C_{\mathrm{completion}}$, $R$, $(C_n)_{n=1}^{N}$ such that
\begin{equation}
\begin{split}
\label{dfabaann}
& P_{m,s_m}C_{\mathrm{completion}}P_{m,s_m} = 0,\quad m = 1,\ldots, M,\quad s_m = 1,\ldots,S_m,\\
& P_{m,s_m}C_{\mathrm{completion}}P_{m',s'_{m'}} = 0,\quad m,m' = 1,\ldots, M,\quad m\neq m',\quad s_m = 1,\ldots,S_m,\quad s'_{m'} = 1,\ldots, S_{m'},\\
\end{split}
\end{equation}
and 
\begin{eqnarray}
& & C  =  C_{\mathrm{observable}} + C_{\mathrm{completion}},\\
\label{qefffadbg} &  & C  =  R + \sum_{n=1}^{N}C_n,\quad R\geq 0,\quad C_n\geq 0,\\
\label{qfqwbgrb}
 & &  P^{(n)}C_n P^{(n)} = C_n,\\
\label{wfmdedafv}
& &  R = \sum_{m,s_m}P_{m,s_m}R P_{m,s_m},
\end{eqnarray}
where $P^{(n)}:= \sum_{m\in\mathcal{C}_n}\sum_{s_m=1}^{S_m}P_{m,s_m}$, and $\mathcal{C}_n$ denotes the children having parent $p_n$ in $G$,
 then there exist operators 
$\overline{R}$ and $(\overline{C}_n)_{n=1}^{N}$ such that
\begin{eqnarray}
\label{sfgnfgnfgn} & & \mathrm{Cov}_{\overline{P}}(\overline{Y}) =  \overline{R} + \sum_{n=1}^{N}\overline{C}_n,\quad \overline{R}\geq 0,\quad \overline{C}_n\geq 0,\\
\label{fgngnmfmm}
& &  \overline{P}^{(n)}\overline{C}_n  \overline{P}^{(n)}= \overline{C}_n,\\
& & \label{fgnfgnsfnae}
  \overline{R} = \sum_{m=1}^{M}\overline{P}_m \overline{R} \overline{P}_m,
\end{eqnarray} 
where $ \overline{P}^{(n)} := \sum_{m\in\mathcal{C}_n} \overline{P}_m$.
\end{Proposition}

\begin{proof}
This proof does to a large extent build on Supplementary Proposition \ref{PropSelection}, and as the first step we shall express $\mathrm{Cov}_{\overline{P}}(\overline{Y})$ in terms of  $C^{(\overline{s})}$ defined in (\ref{fgdsrtnsgfrn}). For this purpose we define 
\begin{equation}
\label{adfbadfb}
\begin{split}
\tilde{R} := &    \sum_{m}\sum_{s_m}q^{(m)}_{s_m}\sum_{x_{m}} \sum_{x'_{m}}  \overline{Y}^{(m)}_{x_{m}}{\overline{Y}^{(m)}_{x'_{m}}}^{\dagger}   \Tr([A^{(m,s_m)}_{x_{m}}\otimes\hat{1}_{/c_m}]\rho)    \Tr([A^{(m,s_m)}_{x'_{m}}\otimes\hat{1}_{/c_m}]\rho)\\
&  -  \sum_{m}\sum_{x_{m}} \sum_{x'_{m}}  \overline{Y}^{(m)}_{x_{m}}{\overline{Y}^{(m)}_{x'_{m}}}^{\dagger}   \Tr([\sum_{s_m}q^{(m)}_{s_m}A^{(m,s_m)}_{x_{m}}\otimes\hat{1}_{/c_m}]\rho)    \Tr([\sum_{s'_m}q^{(m)}_{s'_m}A^{(m,s'_m)}_{x'_{m}}\otimes\hat{1}_{/c_m}]\rho)\\
%%%
= & \sum_{m} \sum_{s_m}q^{(m)}_{s_m} \Big[f^{(m)}_{s_m}  -\sum_{s'_{m}} q^{(m)}_{s'_m}f^{(m)}_{s'_m}\Big] \Big[f^{(m)}_{s_m}  -\sum_{s''_{m}} q^{(m)}_{s''_m}f^{(m)}_{s''_m}\Big]^{\dagger}\geq  0,
\end{split}
\end{equation}
where $f^{(m)}_{s_m} :=  \sum_{x_{m}}  \overline{Y}^{(m)}_{x_{m}}  \Tr([A^{(m,s_m)}_{x_{m}}\otimes\hat{1}_{/c_m}]\rho)$.  
Hence  $\tilde{R}$ is positive semidefinite.
 One can also confirm that $\sum_m \overline{P}_m\tilde{R}\overline{P}_m = \tilde{R}$.
We next define  $F :=\sum_{m}\sum_{s_m} V_{m,s_m}^{\dagger}$ and  
$q(\overline{s}) := q^{(1)}_{s_1}\times \cdots \times q^{(M)}_{s_M}$. 
With $C^{(\overline{s})}$ as defined in (\ref{fgdsrtnsgfrn}) we get 
\begin{equation}
\label{adfbqbaebtb}
\begin{split}
\sum_{\overline{s}}q(\overline{s})FC^{(\overline{s})}F^{\dagger}= &   \sum_{m}\sum_{s_m}q^{(m)}_{s_m}F\mathrm{Cov}(Y^{(m,s_m)})F^{\dagger}  +\sum_{\substack{m,m' \\ m\neq m'}}\sum_{s_m,s_{m'}} q^{(m)}_{s_m}q^{(m')}_{s_{m'}}F\mathrm{Cov}(Y^{(m,s_m)},Y^{(m',s_{m'})})F^{\dagger}\\
= &    \sum_{m}\sum_{s_m}q^{(m)}_{s_m}\sum_{x_{m}} \overline{Y}^{(m)}_{x_{m}}{\overline{Y}^{(m)}_{x_{m}}}^{\dagger} \Tr([A^{(m,s_m)}_{x_{m}}\otimes\hat{1}_{/c_m}]\rho)\\
&  -  \sum_{m}\sum_{s_m}q^{(m)}_{s_m}\sum_{x_{m}} \sum_{x'_{m}}  \overline{Y}^{(m)}_{x_{m}}{\overline{Y}^{(m)}_{x'_{m}}}^{\dagger}   \Tr([A^{(m,s_m)}_{x_{m}}\otimes\hat{1}_{/c_m}]\rho)    \Tr([A^{(m,s_m)}_{x'_{m}}\otimes\hat{1}_{/c_m}]\rho)\\
&  +\sum_{\substack{m,m' \\ m\neq m'}}\sum_{s_m,s_{m'}} q^{(m)}_{s_m}q^{(m')}_{s_{m'}}
 \sum_{x_{m}} \sum_{x'_{m'}} \overline{Y}^{(m)}_{x_{m}}  {\overline{Y}^{(m')}_{x'_{m'}}}^{\dagger}\\
& \times \bigg[\Tr(A^{(m,s_m)}_{x_{m}}\otimes  A^{(m',s_{m'})}_{x'_{m'}} \otimes\hat{1}_{/c_mc_{m'}} \rho)- \Tr([A^{(m,s_m)}_{x_{m}}\otimes\hat{1}_{/c_m}]\rho)\Tr(  [A^{(m',s_{m'})}_{x'_{m'}} \otimes\hat{1}_{/c_m}]\rho)\bigg].
\end{split}
\end{equation}
By combining (\ref{adfbadfb}) with (\ref{adfbqbaebtb}) and (\ref{nhkjsfgngfsn}) we get
\begin{equation}
\label{lnfbeionbproe}
 \mathrm{Cov}_{\overline{P}}(\overline{Y}) =  \tilde{R}+  \sum_{\overline{s}}q(\overline{s})FC^{(\overline{s})}F^{\dagger}.
\end{equation}
Hence, we have expressed $\mathrm{Cov}_{\overline{P}}(\overline{Y})$ in terms of $C^{(\overline{s})}$.
 Next, we observe that since $C_{\mathrm{observable}}$ satisfies 
(\ref{dfabaann}), (\ref{qefffadbg}), (\ref{qfqwbgrb}),  and (\ref{wfmdedafv}), it follows that Supplementary Proposition \ref{PropSelection} is applicable to $C^{(\overline{s})}$ for each $\overline{s} = (s_1,\ldots,s_M)$.
By combining (\ref{lnfbeionbproe}) with the decomposition (\ref{fdbflfgnknm}) in Supplementary Proposition \ref{PropSelection} it follows that 
\begin{equation}
\label{dfnjnbdtrgntrn}
 \mathrm{Cov}_{\overline{P}}(\overline{Y}) =  \overline{R} + \sum_{n=1}^{N}\overline{C}_n,\quad \quad
\overline{R} :=   \tilde{R}+  \sum_{\overline{s}}q(\overline{s})FR^{(\overline{s})}F^{\dagger},\quad\quad \overline{C}_n :=   \sum_{\overline{s}}q(\overline{s})FC_n^{(\overline{s})}F^{\dagger}. 
\end{equation}
Next we wish to show that $\overline{R}$ and $(\overline{C}_n)_{n=1}^{N}$ satisfy all the required properties. 
By Supplementry Proposition \ref{PropSelection} we know that $C^{(\overline{s})}_n\geq 0$, and thus  $\overline{C}_n \geq 0$.
From Supplementary Proposition \ref{PropSelection} we also know that $R^{(\overline{s})}\geq 0$, and thus $\sum_{\overline{s}}q(\overline{s})FR^{(\overline{s})}F^{\dagger}\geq 0$. From (\ref{adfbadfb}) we also know that 
$\tilde{R}\geq 0$. We can thus conclude that  $\overline{R} \geq 0$ and  $\overline{C}_n \geq 0$. 

Next, we  show that $\overline{C}_n$ satisfies (\ref{fgngnmfmm}). 
Recall that $\overline{P}^{(n)} = \sum_{m\in\mathcal{C}_n} \overline{P}_m$, $F = \sum_{m}\sum_{s_m} V_{m,s_m}^{\dagger}$, and moreover that $V_{m,s_m}^{\dagger}V_{m,s_m} = \overline{P}_{m}$  and $V_{m,s_m}V_{m,s_m}^{\dagger} = P_{m,s_m}$.
From this follows that $\overline{P}_mF = F\sum_{s_m}P_{m,s_m}$ and thus 
$\overline{P}^{(n)}F = \sum_{m\in \mathcal{C}_n}  \overline{P}_mF  = F\sum_{m\in \mathcal{C}_n} \sum_{s_m}P_{m,s_m} = FP^{(n)}$. 
By combining this observation with (\ref{fgnfgndnfmd}) it follows that $\overline{P}^{(n)}\overline{C}_n\overline{P}^{(n)} =  \overline{C}_n$, and we can thus conclude that $\overline{C}_n$ satisfies (\ref{fgngnmfmm}).
We finally should show that $\overline{R}$ satisfies (\ref{fgnfgnsfnae}).
For this purpose we note  that $\overline{P}_m F =   F\sum_{s_m}P_{m,s_m}$ 
implies $\overline{P}_m FP^{(\overline{s})}_{\overline{m}} =   F\sum_{s'_m}P_{m,s'_m}P^{(\overline{s})}_{\overline{m}} = F\sum_{s'_m}P_{m,s'_m}
P_{\overline{m},s_{\overline{m}}} = \delta_{m,\overline{m}}FP^{(\overline{s})}_{\overline{m}}$.
By combining this observation with (\ref{dfnjnbdtrgntrn}) and with the previous observation $\sum_m \overline{P}_m\tilde{R}\overline{P}_m = \tilde{R}$,  we obtain  (\ref{fgnfgnsfnae}).
\end{proof}

%%%%%%%%%%%%%%%%%%%%%%%%%%%%%%%%%%%%%%%%%%%%%%%%%%%%%%%%%%%%

\section{\label{SecComparison}  Comparison with the Finner inequality, inflation and entropic bounds}

In Figures 1 and 2 of the main text we compare the results of our method in two distinct scenarios to results obtained with a special case of the Finner inequality (proven to be valid for quantum networks in Ref. \cite{Renou2019}) given by
\begin{equation}
P(x_1,\ldots,x_n) \leq \sqrt{\prod_i P_i(x_i)},
\label{eq:app_finner_indicator}
\end{equation}
valid for distributions compatible with the corresponding network, where $P$ is the complete distribution of the observable variables, and $P_i$ single-party marginals corresponding to the $i$-th party. Incompatibility with the network is established when any possible outcome assignment $x_1,\ldots,x_n$ makes the inequality invalid. Finner inequality, however, in its general form can be stated as
\begin{equation}
E\left(\prod_j f_j \right) \leq \prod_j \left[E(|f_j|^{1/\eta_j})\right]^{\eta_j},
\label{eq:app_finner_general}
\end{equation}
where $f_j$ are generic post-processing functions of each outcome $x_j$, $E$ are expectation values taken over the corresponding parties' distribution, and $\eta_j$ are exponents satisfying $\sum_{j \in \mathcal{C}_n} \eta_j \leq 1$ for all parents $n$. As mentioned in \cite{Renou2019}, the set of allowed exponents is convex and extremal exponents are nonnegative integers and half-integers. For the cases considered in the main text, we have $\eta_i + \eta_j \leq 1$ for all $i \neq j$, which leads to the possibilities $(\eta_1,...,\eta_n) = (1/2,...,1/2)$, or to a single $\eta_j=1$ while all others are $0$, where $\left[E(|f_i|^{1/\eta_i})\right]^{\eta_i}$ for $\eta_i=0$ is to be taken as $\max_{\{x_i | P_i(x_i) \neq 0\}} |f_i(x_i)|$ and is denoted as $\|f_i\|_\infty$. Considering these extremal cases, we are still left with the choice of processing functions $f_j$. To obtain the inequality in Eq.\ \eqref{eq:app_finner_indicator}, we have considered indicator functions $f_j(x) = \delta_{x,x_j}$, where $\delta_{a,b}$ is the Kronecker's delta, given by $\delta_{a,b}=1$, if $a=b$, and zero otherwise. The general case for dichotomic distributions is analyzed below. 

We first consider the case of a single $\eta_j=1$, and the others $0$, in (\ref{eq:app_finner_general}), which yields
\begin{equation}
E\left(\prod_j f_j \right) \leq \left(\sum_{x_j} P_j(x_j)|f_j(x_j)|\right)\,\prod_{i\neq j}|f_i(x^*_i)|,
\label{eq:app_finner_eta1}
\end{equation}
where $x^*_i$ is any outcome that satisfies $|f_i(x^*_i)| = \| f_i\|_\infty$, corresponding to the contribution of terms with $\eta_i = 0$.
In the following we show that  (\ref{eq:app_finner_eta1}) is satisfied trivially.
By the definition of $\| f_i \|_\infty$, either we have $|f_i(x^*_i)| \geq |f_i(x_i)|$ for all $x_i$, or $|f_i(x^*_i)| < |f_i(x^\prime_i)|$ for some $x^\prime_i$, for which $P_i(x^\prime_i) = 0$. If $P_i(x^\prime_i) = 0$, however, then $P(x_1,\ldots,x_n)\vert_{x_i=x^\prime_i} = 0$ as well. Consequently, it is always the case that $P(x_1,\ldots,x_n)\,\prod_{k=1}^n |f_k(x_k)| \leq P(x_1,\ldots,x_n)\,|f_j(x_j)|\,\prod_{i\neq j} |f_i(x^*_i)|$. This inequality yields
\begin{equation}
\begin{split}
 \sum_{x_1,\ldots,x_n}P(x_1,\ldots,x_n)\,\prod_{k=1}^n f_k(x_k) 
\leq & \sum_{x_1,\ldots,x_n}P(x_1,\ldots,x_n)\,\prod_{k=1}^n |f_k(x_k)|\\
 \leq & \sum_{x_1,\ldots,x_n}  P(x_1,\ldots,x_n)\,|f_j(x_j)|\,\prod_{i\neq j} |f_j(x^*_j)|\\
 = &  \sum_{x_j}P_j(x_j)\,|f_j(x_j)|\,\prod_{i\neq j} |f_j(x^*_j)|,
\end{split}
\end{equation}
which implies (\ref{eq:app_finner_eta1}).  Hence,  (\ref{eq:app_finner_eta1}) always holds.

For $\eta_j=1/2$, we need to consider an optimization over possible functions $f_j$. 
Note that, since the right-hand side of (\ref{eq:app_finner_general}) includes only the modulus of each $f_i$, while the left-hand side can become negative with $f_i \leq 0$, we restrict our analysis to $f_i(x_i) \geq 0$ for all parties and values of $x_i$ to look for a violation. In fact, using $f_j \geq 0$, we may square both sides of Eq.\ \eqref{eq:app_finner_general} and obtain
\begin{equation}
\sum_{\substack{x_1,\ldots,x_n\\x^\prime_1,\ldots,x^\prime_n}}\,\left(\prod_j f_j(x_j)\right)\,\left(P(\boldsymbol{x})P(\boldsymbol{x}^\prime) - \prod_j P_j(x_j)\,\delta_{x_j,x^\prime_j}\right)\left(\prod_j f_j(x^\prime_j) \right)\leq 0,
\end{equation}
where $\boldsymbol{x} = (x_1,...,x_n)$ and similarly for $\boldsymbol{x}^\prime$. Determining whether this inequality can be violated can then be interpreted as the constrained quadratic optimization problem
\begin{equation}
\maximize_F \left\{\,F^T\,M\,F \,\left\vert\right. \, F = f_1 \otimes f_2 \otimes \ldots \otimes f_n\right\},
\end{equation}
where $M$ is the matrix with elements $M_{\boldsymbol{x}\boldsymbol{x}^\prime} = P(\boldsymbol{x})P(\boldsymbol{x}^\prime) - \prod_j P_j(x_j)\,\delta_{x_j,x^\prime_j}$. The restriction of $F$ to a pure-product vector makes finding a direct solution to the problem a complicated task numerically, and heuristic methods to find lower bounds to the global maximum of the expression must be used. In particular, note that we may restrict the codomain of $f_j$ to the interval $[-1,1]$ by dividing the inequalities by $\prod_j \sum_{x_j} |f_j(x_j)|$, assuming of course that $f_j(x_j) \neq 0$ for some $x_j$. Using only nonnegative $f_j$, this also implies that $\sum_{x_j} f_j(x_j) = 1$, which, for distributions with only two possible outcomes per party, reduces the problem to optimizing a single variable per party: $f_j(x_j) = (1 + (-1)^{x_j}\,\delta_j)/2$, with $\delta_j\,\in\,[-1,1]$, assuming also $x_j \in \{0,1\}$. By using a sequential optimization for each $\delta_j$, we have obtained the curve included in Figs. 1 and 2 of the main text, both very close to the corresponding curves obtained with the covariance method described in Theorem 1. The algorithm used is described in the following.

For each distribution $P$, we start with a random selection of values for each $\delta_j$, sampled independently from a uniform distribution in the allowed interval $[-1,1]$. We then optimize $\delta_i$ individually, starting with a random $i \in \{1,\ldots,n\}$, where $n$ is the number of parties, and changing according to $i \rightarrow i (\mod\, n) + 1$ for each subsequent step. The result is then combined with the remaining non-updated values of $\delta_j$ for $j \neq i$, to be used as a starting point in the individual optimization. The value of $i$ is updated a fixed number of times and the minimum obtained for $F^T\,M\,F$ is stored. The whole procedure is also repeated a fixed number of times, with the overall minimum collected in the end. To establish the transition graph between non-violating and violating values of $(p,q)$ for the distribution $P^{N}_{pq}(x_1,..,x_N) = p\,\delta^{(N)}_{0} + q\,\delta^{(N)}_{1} + (1-p-q)\,\frac{1 - \delta^{(N)}_{0} - \delta^{(N)}_{1}}{2^N -2}$, we then apply the algorithm to each value of $(p,q)$ evaluated.

Fig. 1 in the main text also includes a comparison with the inflation bound recently obtained in \cite{gisin2020constraints} and the entropic bound \cite{henson2014theory,Chaves2015a}. The three observable variables in the triangle scenario are labelled as $A$, $B$, $C$. The inflation bound was obtained in \cite{gisin2020constraints} by considering the non-signalling constraints arising from a specific inflation of the triangle network and is given by
\begin{equation}
(1 + 2E_1 + E_2)^2 \leq 2(1 + E_1)^3,
\end{equation}
where $E_1=E_A = E_B =E_C$ and $E_2=E_{AB} = E_{BC} = E_{AC}$, with $E_A$, $E_B$ and $E_C$ are the expectation values for single-party marginals, and $E_{AB}$, $E_{BC}$ and $E_{AC}$ are the expectation values for two-party marginals.

In turn, the entropic bound is given by following inequality (and its two other symmetries by relabelling of the parties)
\begin{equation}
I(A:B)+I(A:C) \leq H(A),
\end{equation}
where $H(A)=-\sum_{a} p(a) \log_2 p(a)$ is the Shannon entropy of the outcome measurement $A$ and $I(A:B)= H(A)+H(B)-H(A,B)$ is the mutual information between variables $A$ and $B$.

\bibliography{biblio}